\documentclass[acmlarge]{acmart}

\AtBeginDocument{%
 }

\setcopyright{acmcopyright}
\copyrightyear{2018}
\acmYear{2018}
\acmDOI{XXXXXXX.XXXXXXX}

\acmJournal{POMACS}
\acmVolume{37}
\acmNumber{4}
\acmArticle{111}
\acmMonth{8}

\usepackage{subcaption}

\usepackage{enumitem}
\usepackage{multirow}
\usepackage{tabularx}
\usepackage[resetlabels,labeled]{multibib}
\newcites{SLR}{Survey References}

\usepackage{adjustbox}
\usepackage{array}
\usepackage{booktabs}
\usepackage{setspace}

\newcolumntype{R}[2]{%
 >{\adjustbox{angle=#1,lap=\width-(#2)}\bgroup}%
 l%
 <{\egroup}%
}
\newcommand*\rot{\multicolumn{1}{R{90}{0em}|}}
\newcolumntype{P}[1]{>{\centering\arraybackslash}p{#1}}
\newcolumntype{M}[1]{>{\centering\arraybackslash}m{#1}}

\begin{document}


\title{A Survey on Green Computing in Video Games: The Dawn of Green Video Games}


\author{Carlos P\'erez}
\email{cperez@usj.es}
\orcid{0009-0008-9786-656X}
\affiliation{%
 \institution{Universidad San Jorge}
 \streetaddress{Campus Universitario, Autovía Mudéjar, km. 299}
 \city{Villanueva de Gállego}
 \state{Zaragoza}
 \country{Spain}
 \postcode{50830}
}
\affiliation{%
 \institution{Universitat Polit\`ecnica de Val\`encia}
 \streetaddress{Cam\'i de Vera, s/n}
 \city{Valencia}
 \country{Spain}
 \postcode{46022}
}

\author{Ana C. Marc\'en}
\email{acmarcen@usj.es}
\orcid{0000-0002-5054-4618}
\affiliation{%
 \institution{Universidad San Jorge}
 \streetaddress{Campus Universitario, Autovía Mudéjar, km. 299}
 \city{Villanueva de Gállego}
 \state{Zaragoza}
 \country{Spain}
 \postcode{50830}
}

\author{Javier Ver\'on}
\email{jveron@usj.es}
\orcid{0000-0003-2516-2105}
\affiliation{%
 \institution{Universidad San Jorge}
 \streetaddress{Campus Universitario, Autovía Mudéjar, km. 299}
 \city{Villanueva de Gállego}
 \state{Zaragoza}
 \country{Spain}
 \postcode{50830}
}
\affiliation{%
 \institution{Universitat Polit\`ecnica de Val\`encia}
 \streetaddress{Cam\'i de Vera, s/n}
 \city{Valencia}
 \country{Spain}
 \postcode{46022}
}

\author{Carlos Cetina}
\email{ccetina@usj.es}
\orcid{0000-0001-8542-5515}
\affiliation{%
 \institution{Universidad San Jorge}
 \streetaddress{Campus Universitario, Autovía Mudéjar, km. 299}
 \city{Villanueva de Gállego}
 \state{Zaragoza}
 \country{Spain}
 \postcode{50830}
}
\renewcommand{\shortauthors}{P\'erez et al.}

\begin{abstract}
    
 Today, the large number of players and the high computational requirements of video games have motivated research on Green Video Games. We present a survey that provides an overview of this recent research area. A total of 2,637 papers were reviewed, selecting 69 papers as primary studies for further analysis. Through a detailed analysis of the results, we propose a new way to define the Green Video Game issues based on motivation, device, and layer of the primary studies. Then, we analyze the different applied techniques, the limitations and levels of evidence, and specific aspects of video games.

\end{abstract}




\maketitle

\section{Introduction}

Video games are no longer a niche hobby but a mass phenomenon. Today, the number of video game players is estimated to be approximately three billion worldwide \cite{ExplodingTopics2023} \cite{Statista2023}. This figure is expected to continue to grow considerably in the coming years.

Video game players use everything from game consoles that are created specifically for playing video games to computers or even mobiles. These computers or mobiles are usually high-performance devices and are known as gaming computers or mobiles. In addition, it is very common for video games to also need a network infrastructure and servers to get from the popular online games to the latest cloud gaming platforms such as Xbox Cloud Gaming \cite{xbox} or Geforce Now \cite{geforce}.

The combination of the large number of players (i.e., users) with the high computational requirements of modern video games has motivated recent research work in the field of what we call Green Video Games. We consider Green Video Games to be the specialization of Green Computing that addresses the peculiarities of video games. Therefore, Green Video Games are video games that are designed, implemented, and made available to players in a way that limits the harmful impact on the environment.

This idea of Green Video Games has also recently gained prominence in the industry. In early 2023, Microsoft released an update to its Xbox game console that, in Microsoft's own words, made it "the first Carbon Aware Console"~\cite{xboxFirst}. At the moment, carbon awareness capabilities are limited to scheduling game, app, and operating system updates based on local carbon intensity data. However, carbon intensity data is potentially beneficial for game developers to develop Green Video Games.

To date, surveys have not addressed the sustainability of video games per se. The work in \cite{gutierrez2023green} highlights that there are two perspectives for analyzing energy consumption: Green-IN and Green-BY. The first one deals with improving the energy consumption of a field per se, e.g., video games, whereas the second one is focused on the application of a field (video games) to improve sustainability in any context. It is possible to find some surveys on GREEN-BY video games. For example, \cite{johnson2017gamification} conducted a systematic review to assess the effectiveness of gamification and serious games in impacting domestic energy consumption. However, we did not find any surveys on GREEN-IN video games, which is the perspective of this work.

Our paper is the first to study the current state of Green Video Games. Based on the analysis of 69 papers published between 1995 and 2022, this paper reports on the work done in Green Video Games. In the first part of our review, we propose a way to classify current and future work on Green Video Games. Then, the survey results provide an up-to-date, holistic review of the field of Green Video Games. This article also aims to serve to identify recommendations for practitioners, outstanding research challenges, and several potential avenues of future research for Green Video Games. 

Specifically, our survey reveals that this first wave of Green Video Game work includes work that leverages energy-efficient techniques from the hardware and networking communities for video games, e.g., dynamically distributing computational tasks between mobile devices and servers (resource allocation) or adapting CPU or GPU frequency to the tasks being performed (frequency scaling). Much of this work is motivated by cost, user satisfaction, and user experience. For example, users may not be satisfied if the battery of their mobiles is rapidly consumed by a video game. However, it is an opportunity for the community to further develop video games to have a minimal environmental impact.

There are also efforts on the software side to improve efficiency in basic algorithms for video games such as pathfinding, tree search, or image anti-aliasing. We have not found any explanation of why there is a lack of effort on essential artifacts in video game development such as engines in the literature. Very few papers address physics engines, and only two works address a proprietary engine. No paper addresses either the most popular game engines such as Unreal and Unity or even sets of essential physics, graphics, or sound libraries (e.g., ogre3d or bullet). We can only theorize (from the trajectories of the researchers) that, at this early stage, the Green Video Game community is probably made up of "green researchers" who have approached video games, rather than video game researchers who have approached the green aspect.

Instead of exploring a top-down strategy where techniques focus on the art and design layers of video games, most studies explore a bottom-up strategy where techniques focus on the hardware, network, or software layers. However, design artifacts (e.g., of corridors or the location of key elements) can be useful to answer questions such as what the influence of the game engine on energy consumption is. Even the influence of the engine per content type can be studied, since in the field of video games the content types (NPC, textures, maps, sounds, etc.) are almost like subfields. In all cases, artifacts in the top layers are essential to video games and are underexplored from a green perspective, which is an opportunity.

Finally, as far as video game genres are concerned, the genre that has received the most attention is Massively Multiplayer Online Games (MMOG). We have found no reason for other genres to be neglected, so it is an opportunity for the community to explore other popular genres such as First-Person Shooter (FPS) or Real-Time Strategy (RTS), especially when genres like these are also among the most played nowadays.

The remainder of this paper is structured as follows: Section ~\ref{section:background} describes some concepts and related works related to the survey. Section~\ref{section:search} describes the research method. Section~\ref{section:results} presents the results of the survey performed according to the research questions. Section~\ref{section:discussion} and Section~\ref{section:threats} discuss the results and the threats to validity, respectively. Finally, Section~\ref{section:conclusions} summarizes the main conclusions.

\section{Background}
\label{section:background}

We consider Green Video Games as the specialization of Green Computing that addresses the peculiarities of video games. Therefore, to better understand Green Video Games, this section briefly describes some concepts of Green Computing. This section also serves to describe the existing surveys that are related to the current research field.

\subsection{Green Computing}

Green Computing is responsible for designing, manufacturing, using, and disposing of computers, servers, and hardware such as monitors, printers, storage devices, and networking and communications systems in order to efficiently and effectively consume energy with minimal or no impact on the environment \cite{chandrakant2013green}. There are five main areas in Green Computing \cite{dhaini2021green}:

\begin{itemize}

 \item \textbf{Green Cloud Computing} refers to the environmental benefits that cloud-based services can offer, for example, saving energy by using cloud computational resources dynamically on demand, as virtual machines.

 \item \textbf{Mobile Computing} refers to the energy-efficiently of mobile devices that allow access to information from anywhere, at any time.

 \item \textbf{Green Software} refers to software that is developed to limit energy consumption and have minimal environmental impact.
 
 \item \textbf{Green Data Centers} refers to the design, protocols, equipment, infrastructure, and algorithms that make data centers energy-aware and efficient, and minimize carbon emissions.

 \item \textbf{Educational Sector} refers to the importance of introducing the principles of green computing, and making both students and practitioners aware of its importance.
 
\end{itemize}

There is no specific area for video games studies. The works on video game are distributed in these five areas. For example, \citeSLR{Nery201997} seeks to reduce the energy consumption of pathfinding algorithms, whi ch corresponds to the Green Software area. The work in \citeSLR{han2020virtual} proposes a distributed algorithm to optimize virtual machine placement through resource competition, which corresponds to the Green Cloud Computing area. The work in \citeSLR{Muhuri20182311} focuses on increasing the battery life of mobile devices to improve user satisfaction, which corresponds to the Mobile Computing area. The work in \citeSLR{Behiya202084} evaluates the influence of controlling the virtual machine central process unit (vCPU) on the energy consumption of data centers, which corresponds to the Green Data Centers area. The work in \cite{casals2017promoting} develops a serious game to make people aware of energy consumption and carbon emission reduction, which corresponds to the Educational Sector. 

Therefore, if researchers wanted to obtain an overview of video games in green computing, they would have to study all of the previously mentioned areas. However, this work can help researchers obtain a deeper perspective on Green Video Games by considering Green Video Games as a new area of green computing that has emerged and grown rapidly in recent years.

\subsection{Related Work}

After a careful review of the related literature, we were unable to find other surveys that put the focus on Green Video Games. The closest works related to this one can be found in surveys studying how to increase awareness towards a more sustainable future through serious games.

Serious games are video games whose primary objective is to teach rather than to entertain. For this reason, there are many works that take advantage of serious games to raise energy awareness in an attractive and relaxed way. In fact, it is possible to find several surveys that review the related literature. In \cite{johnson2017gamification}, a systematic review was conducted to assess empirical support for the effectiveness of gamification and serious games in impacting domestic energy consumption. The authors provide varying degrees of evidence of positive influence found for behaviour, cognitions, knowledge and learning, and the user experience. The work in \cite{katsaliaki2012survey} conducts a review of serious games on sustainable development. Among its objectives are to improve the feature-set of these games and to increase knowledge around sustainable development strategies. The work in \cite{wu2020serious} reviews serious games as an engaging medium for building energy consumption. The investigation specifically focuses on the potential impact that serious games can have on changing the domestic practices of householders, in a safe, fun, and interactive environment. The review conducted in \cite{morganti2017gaming} shows that serious games and gamification have been used in three different areas related to energy efficiency: environmental education, consumption awareness, and pro-environmental behaviours. 

Therefore, some surveys focus on providing an overview of how green education benefits from serious games. However, our survey focuses on the opposite perspective: how video games benefit from green approaches or techniques.

In addition, some surveys mention energy consumption in video games. The work in \cite{chalal2022visualisation} conducted a systematic review of a wide range of energy eco-feedback visualization techniques. One of the categories used to report the techniques is game-based visualization. In \cite{sorrell2020digitalisation}, the review links the specific factors that determine the environmental impacts of information and communication technologies to the magnitude of the impacts on energy consumption and carbon emissions. One of the reviewed categories is e-videos and games. However, the studies in this category present a rather mixed picture and suggest less potential for energy saving than the other categories.

Neither of these two articles delves into the ecological perspective of video games or offers a global vision of Green Video Games. Nevertheless, these two articles that were published in the last four years consider video games as a specific category. This may be indicative of the growth of video games in Green Computing. 

\subsection{Research Method}
\label{section:search}

This survey was conducted using formal and reproducible steps to identify, evaluate, and interpret the scientific studies that are related to Green Computing in the development of video games. Specifically, the research method used follows the best practices and guidelines for systematic literature review research~\cite{kitchenham2007guidelines,kitchenham2013systematic,kitchenham2015evidence}. This section describes the review method and how it was applied to address the research questions raised.

\subsection{Research questions}

 Research questions (RQs) are motivated by the need to provide an overview of the Green Video Game area. This survey aims to answer the following RQs:
 
\begin{enumerate}[label=RQ\arabic*:,font=\bfseries\itshape]
 \item \textit{What issues are addressed in the field of Green Video Games?}

 In Green Computing, some works have focused on identifying relevant green issues and evaluating different approaches to these problems \cite{li2011survey,jain2020novel}. The main goal of RQ1 is to identify which of these issues are addressed in the Green Video Game area. In addition, RQ1 also serves to adapt existing Green Computing taxonomies to cover video game issues or to create a new taxonomy for Green Video Games.
 
 \item \textit{Which green techniques are applied for video game development?}

 The main goal of RQ2 is to identify which green techniques are used to design, develop, implement, and make available video games. This will provide an overview of the issues that are addressed in Green Video Games and the techniques that are used to solve them. RQ2 also serves to identify which techniques can potentially be used for solving other types of issues in Green Video Games.
 
 \item \textit{What are the limitations of the current Green Video Game studies?}

 RQ3 is motivated by the interest in identifying future lines of work. Green Video Games have been a rapidly growing research area in recent years. Therefore, researchers and practitioners would benefit greatly from an overview, especially regarding current problems and future research directions. The main goal of RQ3 is to identify open challenges and possible future research directions for improvement. RQ3 also serves to learn about the maturity of studies published taking into account the level of evidence provided for the current works in the research area.
 
 \item \textit{What other video game aspects should be considered in developing Green Video Games?}
Finally, there are specific aspects of video games that are largely neglected in Green Computing, for example, the genre of video games. A single-player offline game does not require the same infrastructure and energy as a MMOG. For this reason, the main goal of RQ4 is to evaluate other video game aspects that should be considered in order to improve the sustainability of video games. RQ4 also serves to open future discussion lines about Green Video Games.
 
\end{enumerate}

\subsection{Search Strategy}

A database search was adopted to collect the available published literature that is relevant to the RQs. Specifically, we formulated our search strategy based on the guidelines provided in ~\cite{kitchenham2007guidelines,kitchenham2013systematic,kitchenham2015evidence}.
The strategy was composed of the following five steps:

\begin{enumerate}
\item A \textit{search string} was used to collect the primary studies present in the \textit{digital sources}. We found 2,637 articles through the search string.

\item The entire list of retrieved articles was filtered to exclude non-relevant articles. This step was conducted by one author of this paper, who applied the \textit{exclusion criteria} considering the title, keywords, and abstract of the retrieved articles. As a result, 1,020 articles were discarded because they met at least one of the exclusion criteria. The remaining 1,617 articles were selected for the next step.

\item The articles that were selected in the previous step were filtered to include only the relevant articles to answer the RQs. This step was conducted by two authors of this paper, who applied the \textit{inclusion criteria} considering the abstract and full text of the selected articles. As a result, 47 articles were included because they met all of the inclusion criteria, which was about 3\% of the papers found in the first step.

\item To extend the scope of research and to minimize the risk of missing relevant studies, we applied snowballing to identify additional sources~\cite{wohlin2014guidelines}. Specifically, we conducted \textit{backward snowballing} of all of the studies included in the previous step. Then, two authors reapplied the exclusion/inclusion criteria based on the title, abstract, and keywords of all of the new articles found (i.e., 913 articles). As a result, 22 articles were selected for further analysis.

\item In the last step, we performed the \textit{data collection} from the primary studies. This step was not used for filtering purposes, so none of the primary studies were included or excluded in this step. Thus, our survey is based on 69 papers, 47 from the initial query and 22 stemming from snowballing. These 69 papers constitute the primary studies of our survey.
\end{enumerate}

\begin{figure*}
\centering
\includegraphics[width=.99\textwidth]{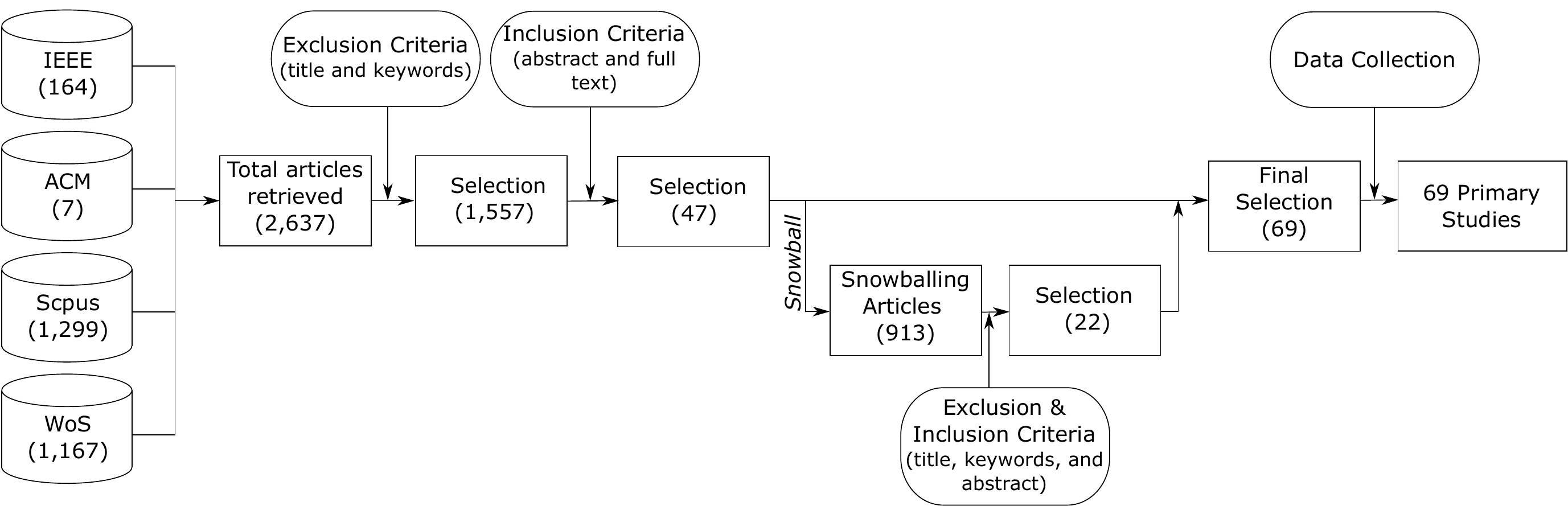}
\caption{Overview of the search strategy}
\label{fig:searchProcess}
\end{figure*}

To ensure accuracy, the selection process was double-checked by another author of this paper. In case of disagreement, the authors held a discussion to reach a consensus. As a result, the 69 papers that were selected as final studies were confirmed as such. Fig.~\ref{fig:searchProcess} shows the search strategy and the number of papers resulting from each of its steps. Moreover, the following sections provide more details about the search string, the digital sources, the selection criteria (i.e., exclusion and inclusion), and the data collection.

\subsubsection{Search String}

The terms for the search strings were extracted following the steps suggested by Kitchenham and Charters~\cite{kitchenham2007guidelines}. First, we used PICO (Population, Intervention, Comparison, and Outcomes) criteria to derive the major terms from the research questions. 

\begin{itemize}
 \item Population: In software engineering, the population may refer to a specific software engineering role, category of software engineer, an application area, or an industry group~\cite{kitchenham2007guidelines}. In our work, the population consists of terms in the video game area.
 \item Intervention: In software engineering, intervention refers to a software methodology, tool, technology, or procedure that addresses a specific issue~\cite{kitchenham2007guidelines}. In our work, the intervention corresponds to green computing terms.
 \item Comparison: In software engineering, the comparison is the software engineering methodology, tool, technology, or procedure against which the intervention is being compared~\cite{kitchenham2007guidelines}. In our work, the comparison is not applied. The goal of this survey is to characterize the articles that tackle Green Computing in video games.
 \item Outcomes: In software engineering, the outcomes should relate to factors of importance to practitioners~\cite{kitchenham2007guidelines}. In our work, the outcomes are terms related to practices, solutions, techniques, or methods considering Green Video Games.
\end{itemize}

Taking into account the RQs, the identified search terms were \textit{video games} (RQ1 and RQ2), \textit{Green Computing} (RQ3), and \textit{techniques} (RQ4). These terms were grouped into three sets: 
\begin{itemize}
 \item Video game terms: those search terms directly related to the population (i.e., \textit{video game*}).
 \item Green terms: those search terms directly related to the intervention (i.e., \textit{Green Computing}).
 \item Approach terms: those search terms directly related to the outcomes (i.e., \textit{techniques}).
\end{itemize} 

Then, we found alternative spellings and/or synonyms. In the case of video game terms, we included \textit{computer game*} and \textit{videogame*}. In the case of Green terms, we included \textit{green software}, \textit{energy efficiency}, and \textit{energy consumption}. In the case of Approach terms, we included \textit{practices}.

We then verified the search terms that other relevant reviews used in their search strings. Video game terms were extended based on the search terms that were used in review articles related to video games (\cite{hall2012health,
connolly2012systematic,
boyle2013systematic, stanitsas2019facilitating}). Green terms and Approach terms were extended based on the search terms that were used in review articles related to Green Computing (\cite{salam2016identification, salam2016developing, rashid2015green, stanitsas2019facilitating}). We did not discard any term found in the search strings. Therefore, all of the new terms that were related to this work were included as terms of our search string.

\begin{table*}[ht]
\renewcommand{\arraystretch}{1.2}
\caption{Search terms}
\label{table:terms}
\begin{tabularx}{\textwidth}{XXX}
\toprule
Video game terms & Green terms & Approach terms\\
\midrule
Video game*, Computer game*, Videogame*, Digital games, Digital video games, Virtual games, Serious gam*, Simulation gam*, Game based learning, MMOG, MMORPG, MUD, Online games, Gamification, Green games, Sustainable Development games, Environmental games &
Green computing, Green software, Energy efficiency, Energy consumption, Sustainable software, Green software development, Green IT, Greener software, Green software engineering, Sustainability, Triple bottom line, Education for sustainability, Sustainable management, Sustainable development &
 Technique*, Practice*, Solution*, Method* \\ \bottomrule
\end{tabularx}
\end{table*}

Table~\ref{table:terms} shows the search terms for each one of the sets: Video game terms, Green terms, and Approach terms. From this table, we used boolean operators to construct the search string. Specifically, all Video game terms were combined by using the boolean 'OR' operator; all Green terms were combined by using the boolean 'OR' operator; and all Approach terms were combined by using the boolean 'OR' operator. Then, we combined the Video game terms, the Green terms, and the Approach terms by using the Boolean 'AND' operator, which implies that, for an article to appear in our search, it needs to include one of the terms for Video game, one of the terms for Green Computing, and one of the terms for Approach. The final search string is reported below:

\textit{("Video game*" OR "Computer game*" OR "Videogame*" OR "Digital games" OR "Digital video games" OR "Virtual games" OR "Serious gam*" OR "Simulation gam*" OR "Game based learning" OR "MMOG" OR "MMORPG" OR "MUD" OR "Online games" OR "Gamification" OR "Green games" OR "Sustainable development games" OR "Environmental games") AND ("Green computing" OR "Green software" OR "Energy efficiency" OR "Energy consumption" OR "Sustainable software" OR "Green software development" OR "Green IT" OR "Greener software" OR "Green software engineering" OR "Sustainability" OR "Triple bottom line" OR "Education for sustainability" OR "Sustainable management" OR "Sustainable development") AND ("Technique*" OR "Practice*" OR "Solution*" OR "Method*")}

\subsubsection{Digital Sources}

The digital sources considered in the survey were IEEE, ACM Digital Library, Scopus, and Web of Science. These sources were selected based on the experience reported by Dyba et al.~\cite{dyba2007applying} and Kitchenham and Brereton~\cite{kitchenham2013systematic}. Specifically, these works reported that the use of IEEE and ACM as well as two indexing sources is sufficient~\cite{petersen2015guidelines}.

This review was carried out in 2023. The search for articles in the four sources was conducted in January 2023. Therefore, the resulting articles belong to 2022 or earlier. Table~\ref{table:searchresults} shows the number of search results per source and the number of selected studies at the end of the search process. 

\begin{table}[t]
\centering
\renewcommand{\arraystretch}{1.2}
\caption{Sources and search results}
\label{table:searchresults}
\begin{tabular}{lccc}
\toprule
\textbf{Source} & \textbf{\# Total studies found} & \multicolumn{1}{l|}{\textbf{\# Primary studies}} \\ \toprule
IEEE & 164 & 15 \\ \midrule
ACM & 7 & 0 \\ \midrule
Scopus & 1,299 & 24 \\ \midrule
Web of Science & 1,167 & 8 \\ 
\bottomrule
\end{tabular}
\end{table}

Note that Table~\ref{table:searchresults} does not reflect the duplicity of papers: a paper can be found in several sources. Due to the search process followed in this work, Scopus papers were selected first, discarding duplicate papers from the other sources. The search then continued by orderly selecting WOS papers, IEEE papers, and ACM papers. With regard to the duplicity of papers, 51\% of the 47 primary studies can be found in Scopus, 17\% of the 47 primary studies can be found in Web of Science, 32\% of the 47 primary studies can be found in IEEE, and 0\% of the 47 primary studies can be found in ACM. Hence, although most of the primary studies were found in Scopus, it is also possible to find them in other sources.

\subsection{Inclusion and Exclusion Criteria}

To answer our research questions, the retrieved articles were evaluated according to a series of exclusion and inclusion criteria. The articles that met one of the exclusion criteria were discarded for our review. The following criteria state the conditions for the exclusion of an article:

\begin{itemize}
 \item Articles that are duplicates of other articles.
 \item Articles that are not written in English.
 \item Articles presenting summaries of conferences or editorials.
 \item Articles presenting non-peer-reviewed material.
 \item Articles presenting review articles (e.g., surveys or systematic literature reviews).
\end{itemize}

On the other hand, the inclusion criteria were applied to include only those articles that were focused on Green Video Games. For this reason, the inclusion criteria ensure that the articles belong to the computer science area, that the articles are related to Green Computing, and that the articles involve the use of video games to a greater or lesser extent.

In addition, when reviewing studies on Green Video Games, it is possible to find two different perspectives: Green-IN and Green-BY \cite{gutierrez2023green}. The first one deals with improving the energy consumption of video games per se. For example, \citeSLR{choi2021optimizing} focuses on improving the energy efficiency of game applications running on mobile devices. In contrast, the second one focuses on the application of video games to improve sustainability in any context. For example, \cite{bang2006powerhhouse} presents a computer game called PowerHouse, which uses video games to influence behaviors associated with energy use and promotes an energy-conscious lifestyle among teenagers. Our work focuses on the first perspective (GREEN-IN), so we have reflected this fact in the inclusion criteria.

The following criteria state the conditions for the inclusion of an article:

\begin{itemize}
 \item Articles that are in the fields of computer science or engineering.
 \item Articles that are related to Green Computing.
 \item Articles that are related to video games.
 \item Articles that focus on improving the energy consumption of video games.
\end{itemize}

To include an article, the article must meet all of the inclusion criteria. The articles that met the inclusion criteria are the ones selected for our review (primary studies). The number of excluded and included articles is shown in Fig.~\ref{fig:searchProcess}.

\subsection{Data Extraction}
\label{sec:data_collection}

The data extraction form was devised to collect data from the primary studies. Specifically, two researchers read the full papers in parallel to collect the data. Table~\ref{tab:data_collection} shows the information collected for each primary study.

\begin{table}[t]
\caption{Data Extraction Form}
\centering
\label{tab:data_collection}
\renewcommand{\arraystretch}{1.2}
\begin{tabular}{|l|l|l|}
\hline
\textbf{\#} & \textbf{Field} & \textbf{Research question} \\ \hline
F1 & Author & n/a \\ \hline
F2 & Year & n/a \\ \hline
F3 & Title & n/a \\ \hline
F4 & Venue & n/a \\ \hline
F5 & Keywords & n/a \\ \hline
F6 & Abstract & n/a \\ \hline
F7 & Citation Count & n/a \\ \hline
F8 & Green Computing Issues defined in \cite{li2011survey} & RQ1 \\ \hline
F9 & Green Computing Issues defined in \cite{jain2020novel} & RQ1 \\ \hline
F10 & Green Video Games Motivations & RQ1 \\ \hline
F11 & Green Video Games Devices & RQ1 \\ \hline
F12 & Green Video Games Layers & RQ1 \\ \hline
F13 & Green Techniques & RQ2 \\ \hline
F14 & Level of Evidence & RQ3 \\ \hline
F15 & Limitations & RQ3 \\ \hline
F16 & Artifacts & RQ4 \\ \hline
F17 & Genre of Video Games & RQ4 \\ \hline
F18 & Content of Video Games & RQ4 \\ \hline
\hline
\end{tabular}
\end{table}

The data extraction was divided into five parts:
\begin{enumerate}
\item Demographic information (F1 to F7): For each paper, the meta-information was extracted for both documentation purposes and the identification of the contribution and quality of the reviewed articles. Fig. \ref{Fig_Demografic_1}, Fig. \ref{Fig_Demografic_2}, Fig. \ref{Fig_Demografic_3}, and Fig. \ref{Fig_Demografic_4} show the distributions of the primary studies concerning publication year, publication type, and number of citations. Fig. \ref{Fig_Demografic_1} and Fig. \ref{Fig_Demografic_3} show that Green Computing in video games has gained popularity in the last decade, with a significant increase in the number of studies published in journals from 2014 to 2022.

\begin{figure}[t]
\centering
\begin{minipage}[b]{.48\textwidth}
\centering
\includegraphics[width=1\textwidth]{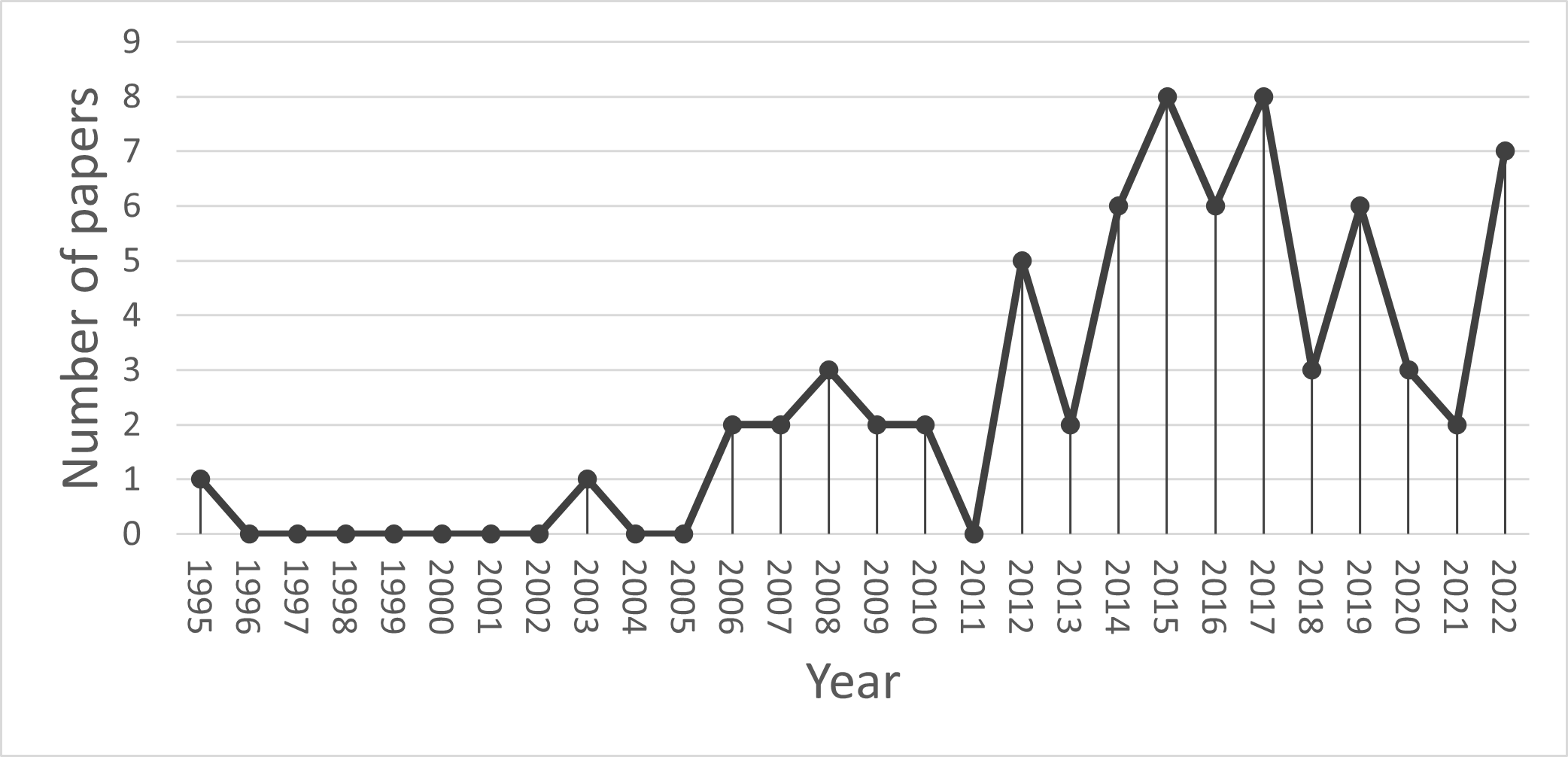}
\caption{Distribution of primary studies over publication year}
\label{Fig_Demografic_1}
\end{minipage}
\hfill
\begin{minipage}[b]{.48\textwidth}
\centering
\includegraphics[width=1\textwidth]{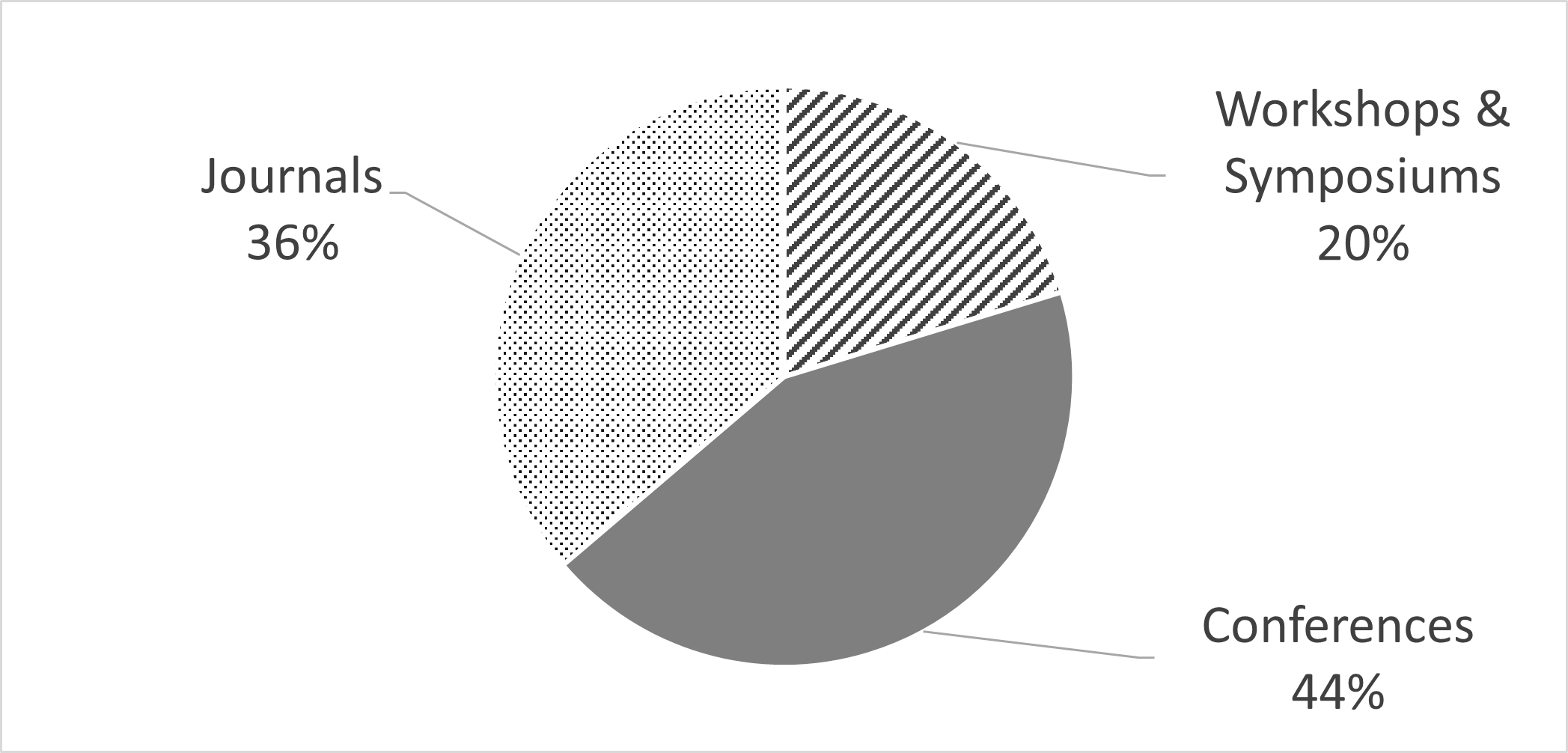}
\caption{Distribution of primary studies over publication type}
\label{Fig_Demografic_2}
\end{minipage}
\begin{minipage}[b]{.48\textwidth}
\centering
\includegraphics[width=1\textwidth]{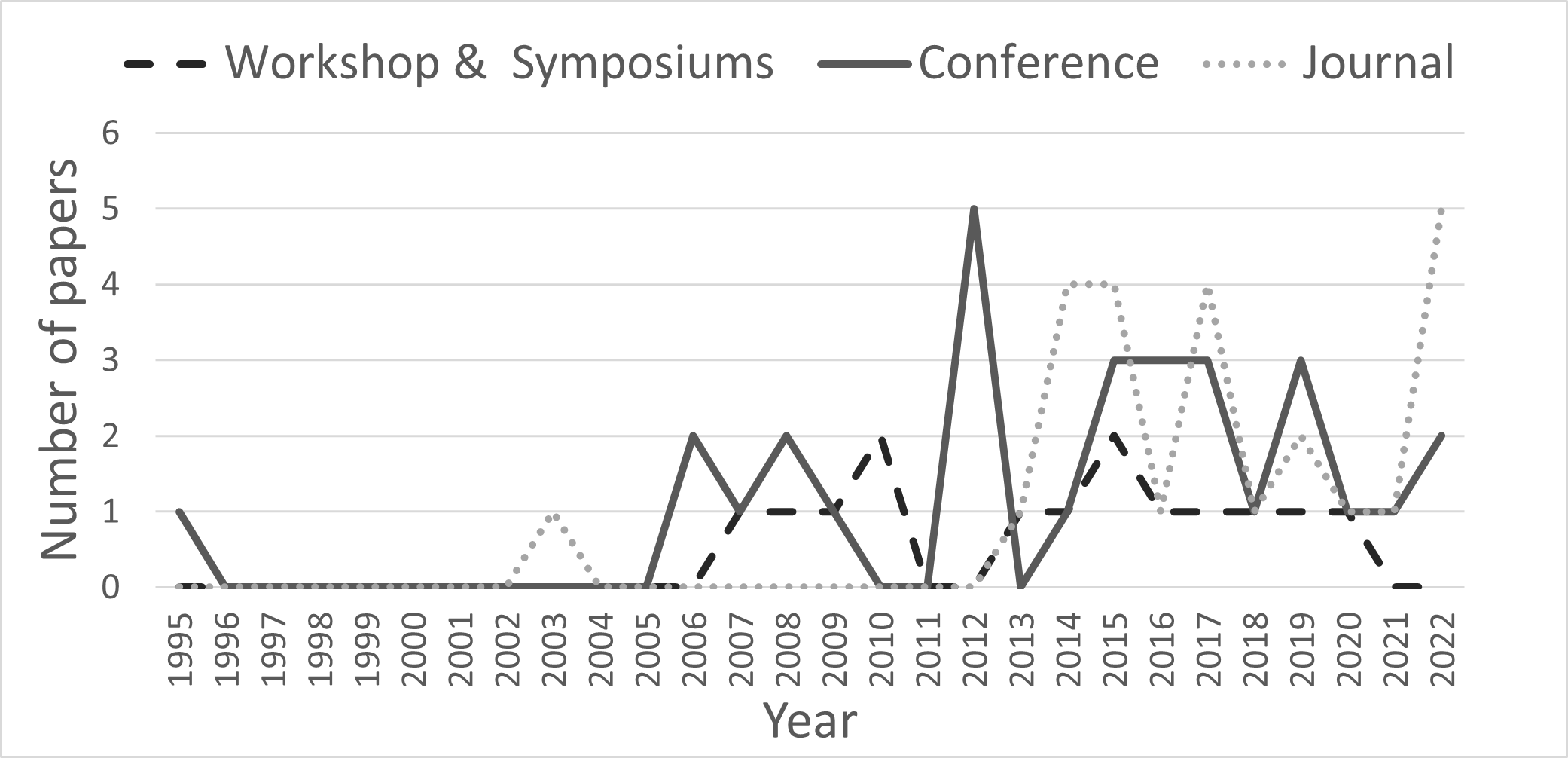}
\caption{Distribution of primary studies over type and year of publication}
\label{Fig_Demografic_3}
\end{minipage}
\hfill
\begin{minipage}[b]{.48\textwidth}
\centering
\includegraphics[width=1\textwidth]{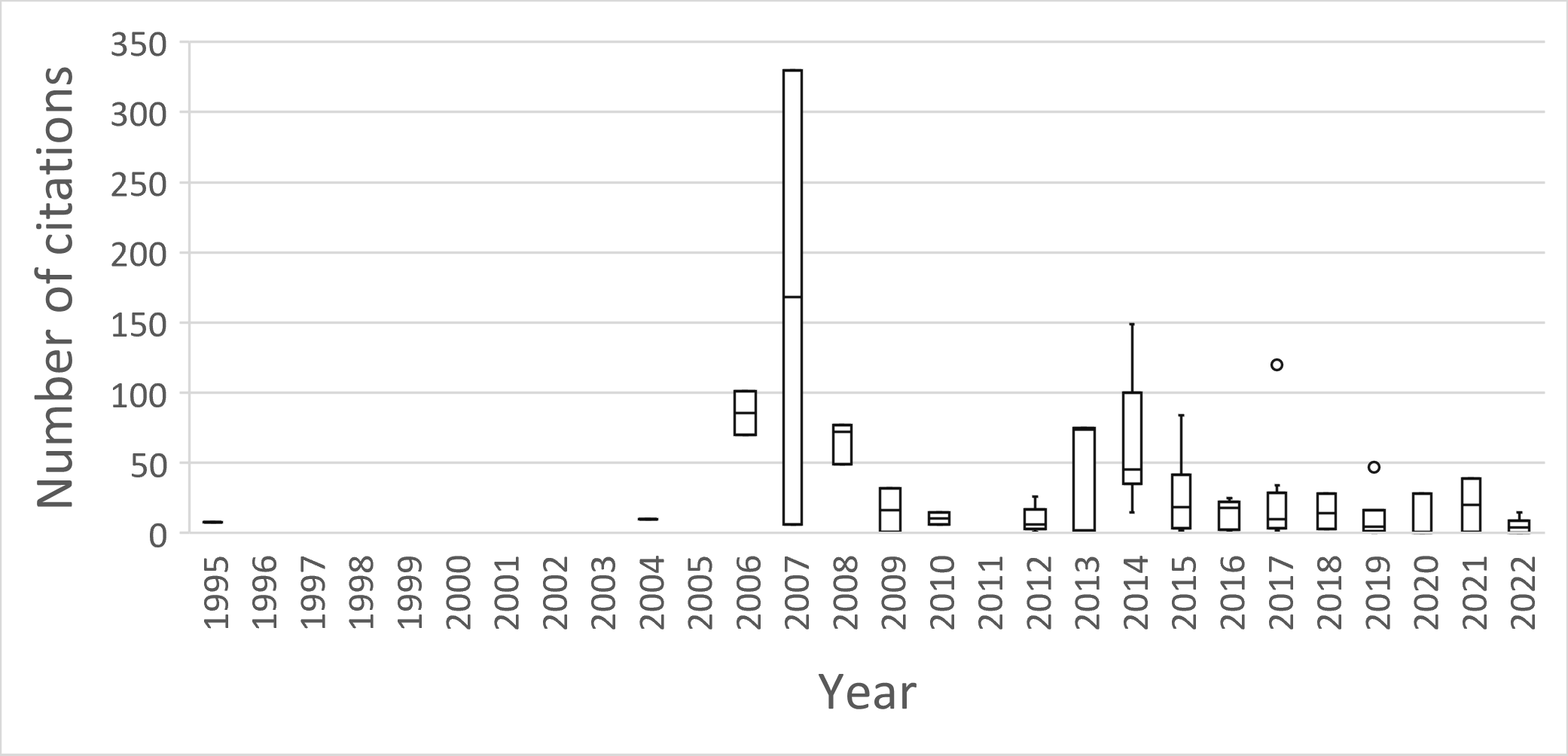}
\caption{Distribution of primary studies over number of citations and year of publication}
\label{Fig_Demografic_4}
\end{minipage}
\end{figure}

\item Data for RQ1 (F8 to F12): For each paper, we extracted the data to answer RQ1. Each of these fields and their categories are detailed below.

F8 considers the taxonomy in \cite{li2011survey} to identify the issue tackled in each reviewed paper. According to this taxonomy, the research articles address three types of issues in Green Computing:
\begin{itemize}
\item \textit{Modeling and evaluation of energy efficiency} focuses on investigating the relationship between energy efficiency and other impact factors, in addition to building the theoretical model and its energy efficiency evaluation mechanism. This problem covers all aspects of computer systems including chips, system architecture, compilers, operating systems, communication networks, applications, etc. 
\item \textit{Energy awareness} focuses on enabling computer systems to run in a low power-consumption manner under a broader range of conditions. Specifically, energy-awareness strategies involve techniques in how systems and applications respond to the state changes and adjust the behaviors of system components. 
\item \textit{Green networking} focuses on all network infrastructure design actions that have an impact on its energy efficiency.
\end{itemize}

F9 considers another taxonomy described in \cite{jain2020novel} to identify the issue tackled in each reviewed paper. This taxonomy is more recent and provides a larger number of issues for Green Computing. Specifically, this taxonomy identifies seven issues for Green Computing:
\begin{itemize}
\item \textit{Product longevity} focuses on extending the lifetime of the equipment, taking into account not only the manufacturing but also its upgradeability and maintainability. 
\item \textit{Data center design} focuses on improving the energy efficiency of data centers, where large amounts of energy are used primarily for information technology systems, environmental conditions, air management, cooling systems, and electrical systems. 
\item \textit{Software and deployment optimization} focuses on algorithmic efficiency as well as resource allocation, terminal servers, and virtualization.
\item \textit{Power management} focuses on different techniques that allow the control of the energy consumed by the components of computer. This includes techniques such as DVFS (Dual voltage Frequency scaling) or the selective disconnection of different components when possible.
\item \textit{Materials recycling} focuses on the recycling of computer equipment, especially those components with potentially toxic or hazardous materials (e.g., lead, mercury, and hexavalent chromium).
\item \textit{Telecommuting} focuses on assessing the energy impact of moving from a face-to-face to a telecommuting work model.
\item \textit{Telecommunication network devices energy indices} focuses on researching the relationship between network technology and environmental impact. Specifically, some studies try to identify the key energy indices that allow a relevant comparison between different devices (network elements).
\end{itemize}

The taxonomies used in F8 and F9 are oriented to Green Computing in general and not to the specific issues of video games. For example, some video games need to render 3D scenes, which requires determining which parts of every object are visible in the final image. This has an important impact on the energy efficiency of a GPU and the memory, 
so there are works that focus on studying the trade-off between energy consumption and application performance, such as \citeSLR{Corbalan-Navarro20224375}. However, the trade-off between energy consumption and application performance is not an issue in the Green Computing taxonomies.

In addition, identifying issues in green video games can be confusing because you need to differentiate between motivation and solution. For example, the motivation of \citeSLR{7351775} considers that the battery in mobile devices is a critical reason to control consumption, so it proposes an energy efficiency evaluation mechanism. Thus, according to its motivation, the study would be classified as a Battery life issue, but according to its solution, it would be classified as a Modelling and measurement of energy consumption issue. 

Therefore, in Green Video Games, some issues cannot be classified using the Green Computing taxonomies and the term \textit{issue} is ambiguous if we do not differentiate between motivation and solution. For these reasons, F10, F11, and F12 were used to collect the main issues tackled from a video game perspective.

F10 considers the motivation. Obviously, some of the motivations are based on the Green Computing issues from the previous taxonomies, but others may be new motivations for the field of Green Video Games.

In contrast, F11 and F12 focus on the solution. Since F11 considers the devices that are used to test the solution (i.e., mobile devices, handheld devices, wearable devices, computers, etc.), it focuses on devices to identify whether or not the proposed solution is device-specific.

Moreover, F12 considers the layers of the proposed solution: \textit{Hardware}, \textit{Network}, \textit{Software}, \textit{Design}, and \textit{Art}. For example, The work in \citeSLR{7351775} proposes a memory-aware cooperative CPU-GPU dynamic voltage and frequency scaling governor. Therefore, this solution affects the hardware layer.

\item Data for RQ2 (F13): For each paper, we extracted the green techniques used to design, develop, implement, and make available video games in order to answer RQ2. To do this, the techniques are collected using natural language.

\item Data for RQ3 (F14 to F15): For each paper, we extracted the data to answer RQ3. Specifically, F14 records the evidence level, which is used to check the maturity of the technique, practice, solution, or method presented in the paper. F14 records the limitations of the primary studies in order to identify open issues and possible future research directions for improvement.

While the data for F14 was collected using natural language, the data for F13 was collected using a scale. We adapted the scale that is described in~\cite{galster2013variability}, which is based on the levels of study design proposed by~\cite{kitchenham2004procedures}. This scale has five levels: (0) no evidence, (1) evidence from demonstration or toy examples, (2) evidence from expert opinions or observations, (3) evidence from academic studies, (4) evidence from industrial studies, and (5) evidence from industrial practice. 

Similarly, our scale has four evidence levels: (0) \textit{no video game}, (1) \textit{toy example or no commercial video game}, (2) \textit{specific video game}, and (3) \textit{multiple video games}. The weakest evidence level of our scale indicates that the paper does not present any evidence (i.e., the presented technique, practice, solution, or method is not evaluated in a video game). Moreover, the strongest evidence level of our scale indicates that the paper presents strong evidence (i.e., the presented technique, practice, solution, or method is evaluated in multiple video games). 

\item Data for RQ4 (F16 to F18): For each paper, we extracted the data to answer RQ4, which considers different aspects of video games that are largely neglected in Green Computing. Specifically, F16 records which artifacts of the video games are considered for the primary studies, such as game engines. F17 records the video game genres, such as MMOG. F18 records which content of the video games is considered to improve sustainability in video games, such as maps, audio, or textures. The data for these fields was collected using natural language.
\end{enumerate}

\section{Results}
\label{section:results}

This section presents the outcomes of the survey based on RQs. The primary studies are listed in the Survey References section, which comes after the bibliography section of this work.

In addition, the results of the whole search process, step by step, are available as supplementary material in \url{https://svit.usj.es/gvg-survey/}. Specifically, this material contains the results of the selection process as well as the results of the snowballing and data collection.

\subsection{RQ1. What issues are addressed in the field of Green Video Games?}

\begin{figure}[t]
\begin{minipage}[b]{.45\textwidth}
\centering
\includegraphics[width=1\textwidth]{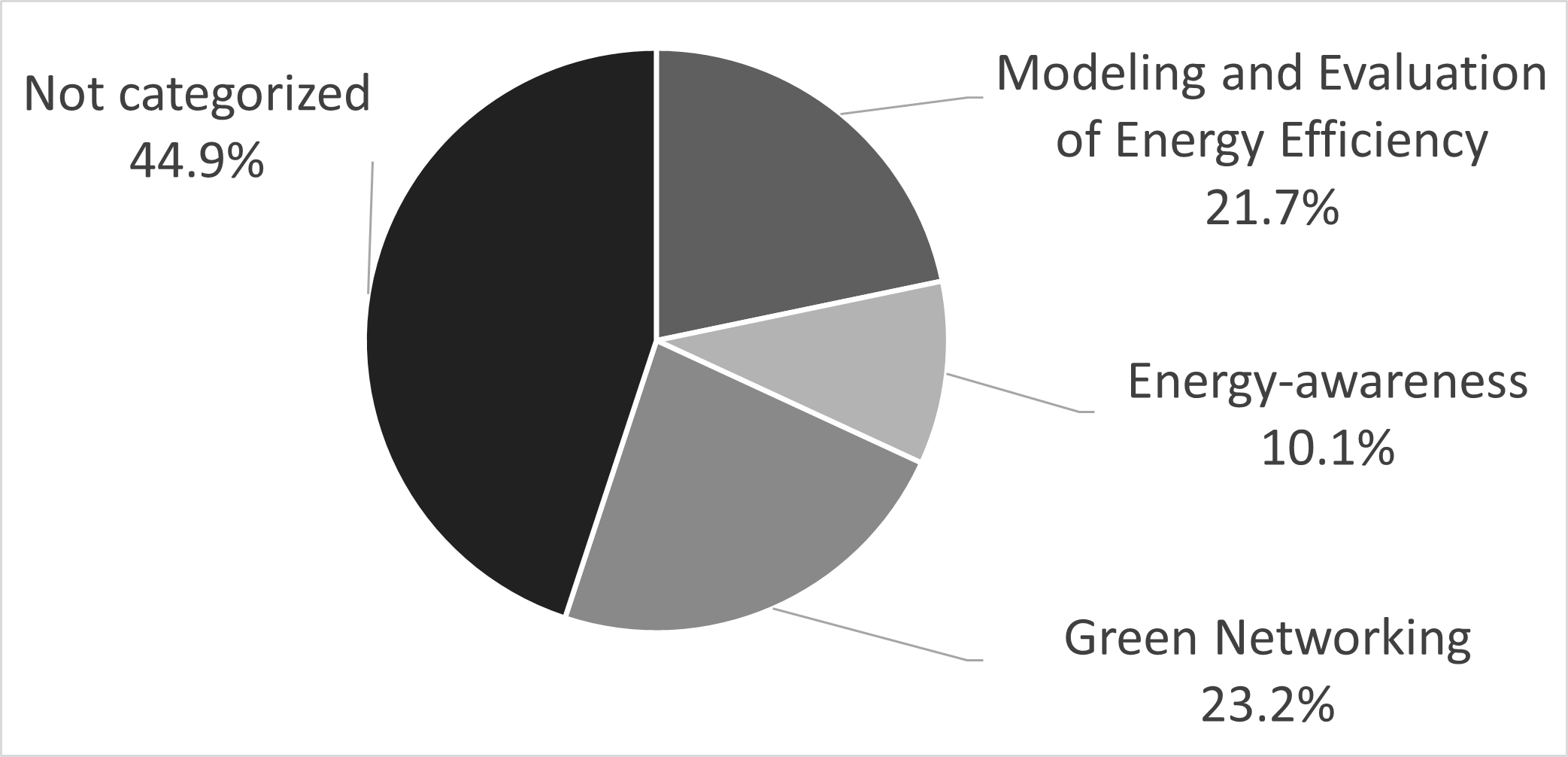}
\caption{Primary studies classified by the Green Computing issues described in \cite{li2011survey}}
\label{Fig_RQ1_1}
\end{minipage}
\hfill
\begin{minipage}[b]{.45\textwidth}
\centering
\includegraphics[width=1\textwidth]{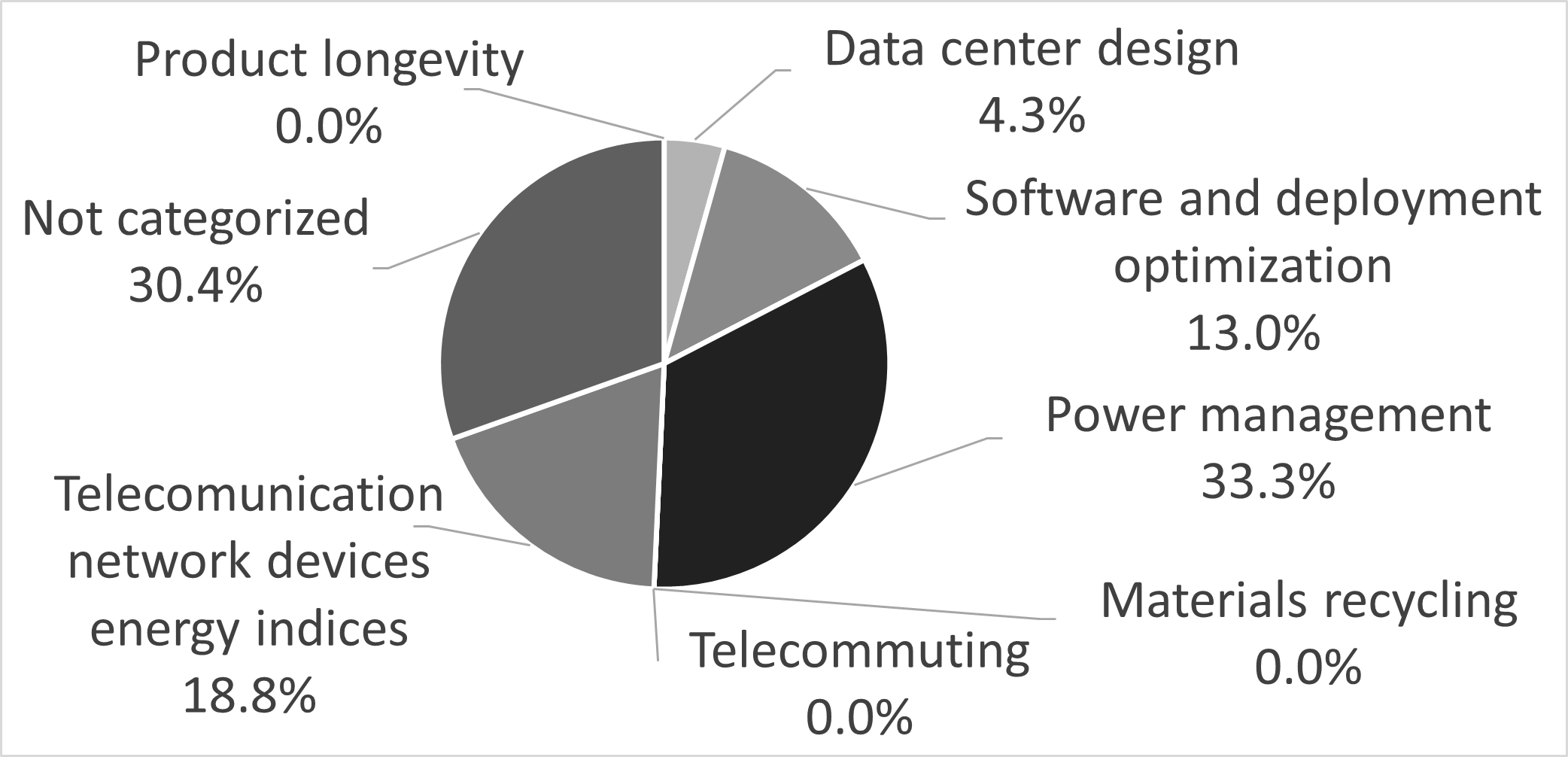}
\caption{Primary studies classified by the Green Computing issues described in \cite{jain2020novel}}
\label{Fig_RQ1_2}
\end{minipage}
\end{figure}

 Following the taxonomy described in \cite{li2011survey}, 15 primary studies focus on the modeling and evaluation of energy efficiency, 7 primary studies focus on energy-awareness, and 16 primary studies focus on green networking. However, 31 primary studies tackled green issues that are not considered by this taxonomy. For this reason, 44,93\% of the primary studies in Fig. \ref{Fig_RQ1_1} are not categorized.

 In contrast, following the taxonomy described in \cite{jain2020novel}, 3 primary studies focus on Data center design, 9 primary studies focus on Software and deployment optimization, 23 primary studies focus on power management, and 13 primary studies focus on Telecommunication network devices energy indices. There are no primary studies that focus on Product longevity, material recycling, and telecommuting. Moreover, there are still 21 primary studies that are not categorized because they tackled green issues that are not considered by the taxonomy.

Fig. \ref{Fig_RQ1_1} and Fig. \ref{Fig_RQ1_2} show which Green Computing issues are addressed in the Video Game area, regarding the issues defined in \cite{li2011survey} and \cite{jain2020novel}, respectively. According to these figures, most of the primary studies focus on power management. However, the taxonomies do not cover all of the green issues that can be encountered in the context of video games. In addition, it could be discussed or assessed whether the categories that have a percentage of 0\% should be considered as future lines of research in this context.

Since the Green Computing issues do not cover all the primary studies, F10 in data extraction allows us to identify which new issues are present in Green Video Games. Specifically, we identified three new issues:
\begin{itemize}
 \item \textit{Battery life:} The rise of mobile devices as a platform for the consumption of video games and the increase in the complexity and graphic resolution of applications has shown that battery life limits or impairs the gaming experience. Therefore, numerous studies address the improvement of the energy efficiency of video games, not so much for environmental concerns but for the impact that lower energy consumption has on battery life and, consequently, on the user's experience.

 \item \textit{Trade-off between energy consumption and application performance, quality of experience, and/or economical cost:} Many techniques for reducing the energy consumption of computer applications, in particular video games, such as Dynamic voltage and frequency scaling (DVFS), can have a negative impact on application performance or the user's experience, and others can have an influence on the economical cost. Many studies are therefore focused on analyzing the ideal combination of parameters to maximize energy savings with minimum degradation of application performance, quality of experience, and acceptable cost.

 \item \textit{Modelling and measurement of energy consumption:} Improving the energy efficiency of the software requires the programmer to know precisely the contribution of each element or process to the overall energy consumption. The level of detail of these studies therefore determines the level at which we can intervene in a controlled manner. Studies that seek to measure the energy consumption of software under controlled conditions and to be able to model its behavior under different conditions from these measurements are therefore a very useful tool in the evolution of software (more specifically of video games towards more energy-efficient versions).
\end{itemize}

Fig. \ref{Fig_RQ1_3} shows the primary studies classified taking into account these new issues. Comparing this figure with Fig. \ref{Fig_RQ1_1} and Fig. \ref{Fig_RQ1_2}, it can be observed that only one issue remains the same: Data center design. The Software and deployment optimization is still an issue, but the number of articles in this category is lower than in \cite{jain2020novel}. This does not mean that the papers are incorrectly classified using the Green Computing issues or that the other issues addressed in Green Computing are not addressed in the context of video games. It means that the strategy used to identify an issue is different in the Video Game area.

\subsubsection{Motivation}
\hfill \break
The issues in Green Video Games revolve more around a specific \textbf{motivation}. For example, considering that gaming applications consume significant power due to the heavy utilization of CPU-GPU resources, \citeSLR{7351775} considers that the battery in mobile devices is a critical reason to control consumption. This is the main motivation of the work. The solution that the authors propose consists of a memory-aware cooperative CPU-GPU dynamic voltage and frequency scaling governor. This governor is an energy efficiency evaluation mechanism (i.e., Modeling and Evaluation of Energy Efficiency according to the Green Computing issues in \cite{li2011survey}) that allows the control of the energy consumed by the CPU-GPU (i.e., Power Management according to the Green Computing issues in \cite{jain2020novel}). Since Green Computing issues do not have a clear difference between motivation and solution, \citeSLR{7351775} would be classified based on the perspective of the researcher. However, in Green Video Games, this may lead to contradictions in classification. Following the above example, one researcher might classify the paper as a \textit{Battery life} issue because he focuses on the motivation, but another researcher might classify the paper as a \textit{Modelling and measurement of energy consumption} issue because he focuses on its solution. Therefore, although the differences between motivation and solution may not matter in Green Computing, in Green Video Games it is important to avoid misunderstandings. 

\begin{figure}[t]
\centering
\begin{minipage}[b]{.45\textwidth}
\centering
\includegraphics[width=1\textwidth]{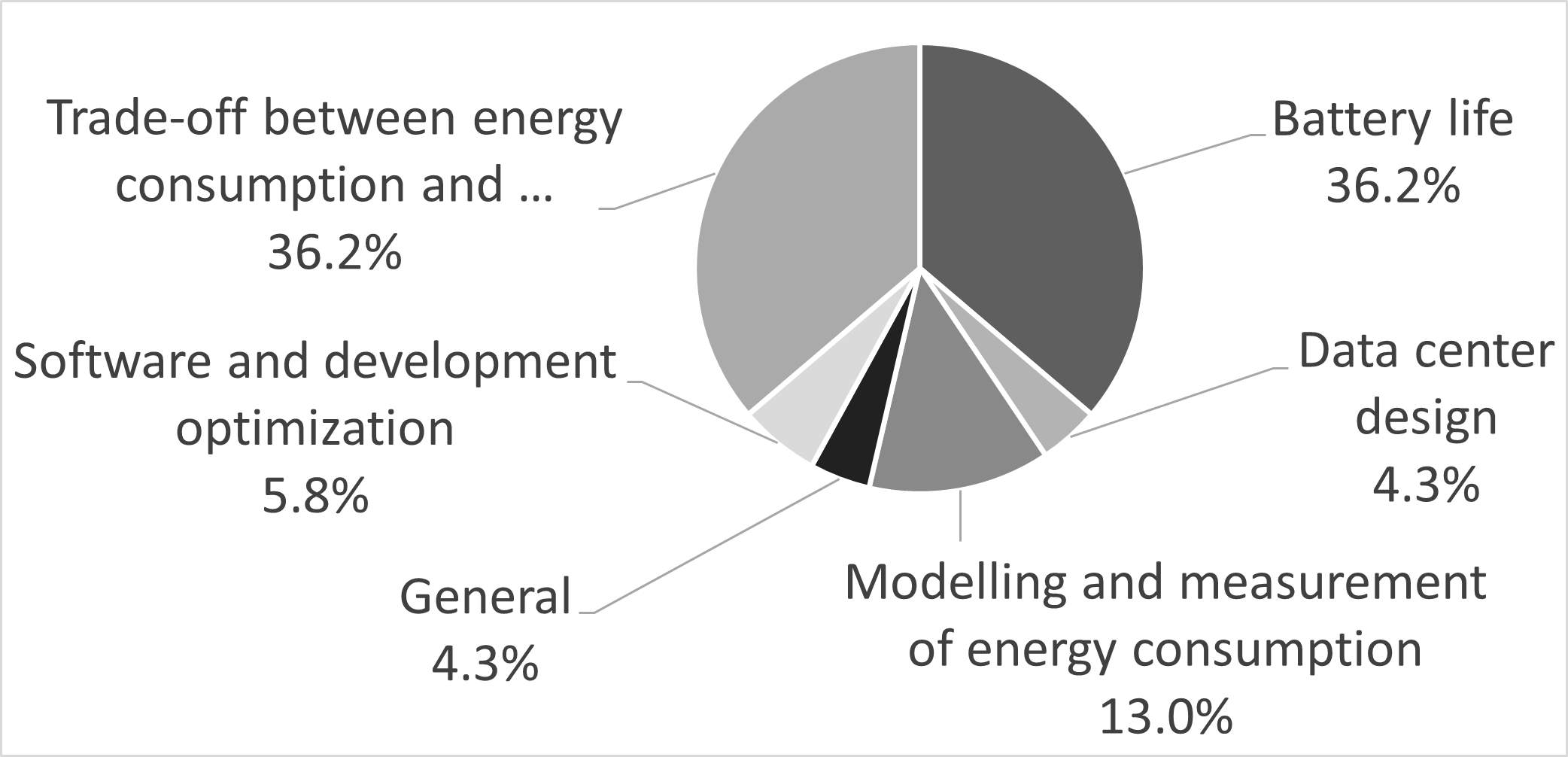}
\caption{Primary studies classified considering motivations in Green Video Games}
\label{Fig_RQ1_3}
\end{minipage}
\hfill
\begin{minipage}[b]{.45\textwidth}
\centering
\includegraphics[width=1\textwidth]{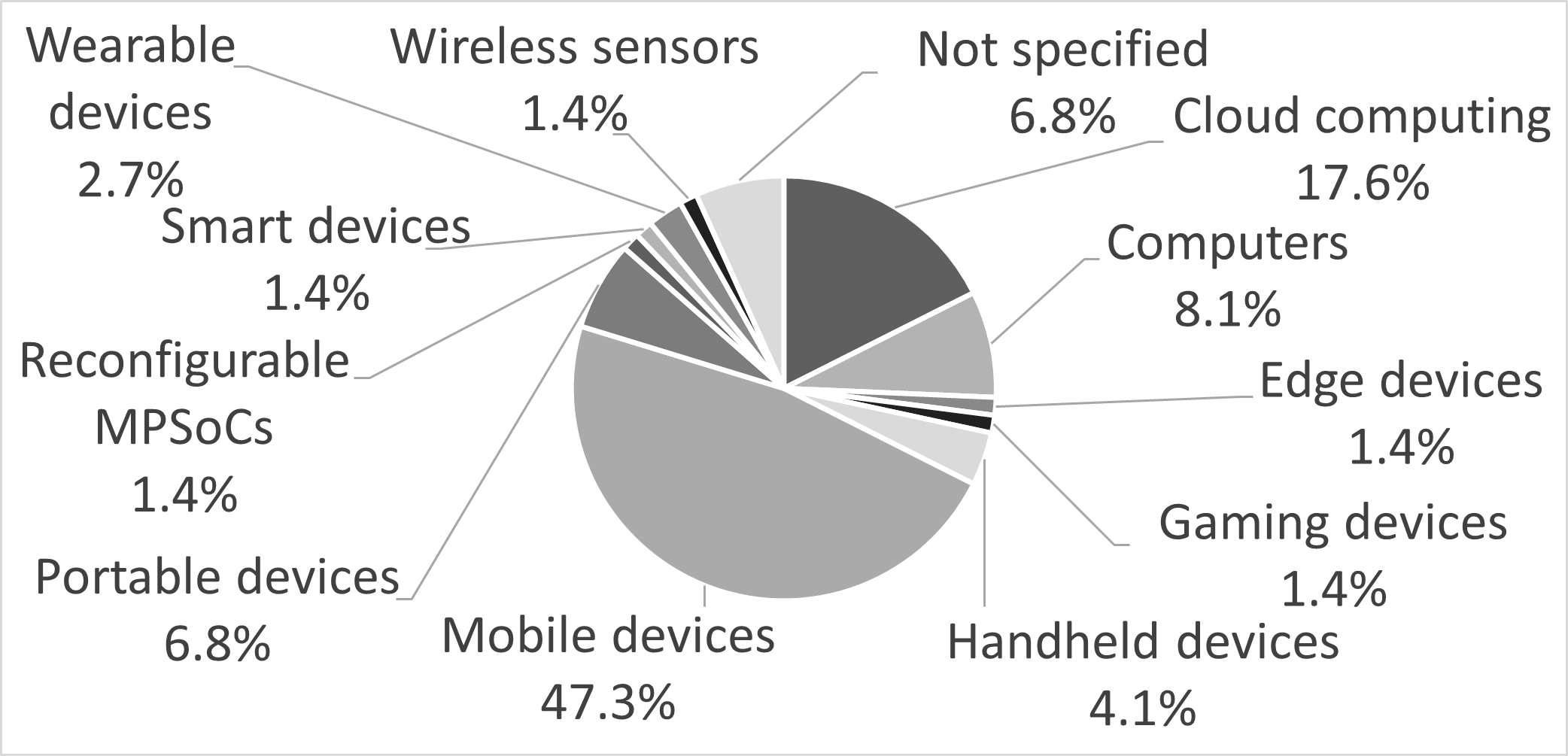}
\caption{Primary studies classified considering devices in Green Video Games}
\label{Fig_RQ1_4}
\end{minipage}
\end{figure}

Fig. \ref{Fig_RQ1_3} shows the papers classified based on motivations. Most of the primary studies tackled Battery life (36,2\% of the primary studies) or the Trade-off between energy consumption and other aspects (36,2\% of the primary studies) as main motivations. The number of primary studies tackling the Modelling and measurement of energy consumption also stands out (13\% of primary studies).

The main studies, whose motivation is the Trade-off between energy consumption and other aspects, compare energy consumption with at least one of these aspects: application performance, quality of experience, and economic cost. Specifically, 20 studies explore the trade-off of energy consumption and application performance, 7 studies explore the trade-off of energy consumption and quality of experience, and 1 study explores the trade-off of energy consumption and economic cost.

Furthermore, three studies are classified as General. These articles review and attempt to classify various aspects of interest related to the environmental impact of video game development. They do not focus on a specific motivation, but rather show an overall view.

\newpage
\subsubsection{Device}
\hfill \break
The issues in Green Computing usually focus on computers. However, Video Games are played on multiple devices, not only computers. Therefore, the issues in Green Video Games depend on devices. For example, video games consume significant power, so the battery in portable devices is critical, but it does not have the same importance in computers. 

Fig. \ref{Fig_RQ1_4} shows the papers classified according to devices. Most of the primary studies (44\% of the primary studies) focus on mobile devices. The number of primary studies focusing on cloud computing also stands out (18\% of primary studies). However, the number of studies that focus on computers is low in Green Video Games (8\% of primary studies).

\subsubsection{Layer}
\hfill \break
The issues in Green Computing are defined providing great importance to the infrastructure layer (i.e., hardware, network, software, etc.). For example, \textit{Power management} focuses on different techniques that allow the control of the energy consumed by the components of computers. This definition is clearly referencing to hardware in computers. However, in Green Video Games, articles define the issues independently of the infrastructure layer. 

For example, Battery life is the motivation for \citeSLR{pathania2014integrated}, \citeSLR{9816214}, \citeSLR{4382127}, and \citeSLR{hosseini2012energy}. However, these three studies focus on a different layer. The work in \citeSLR{pathania2014integrated} proposes reducing the power consumption using a unified view of a CPU-GPU Dynamic Voltage Frequency Scaling (i.e., Hardware). The work in \citeSLR{9816214} proposes an adaptive bitrate model to reduce the transfer latency that allows downloading massive data from the server keeping the user perception experience (i.e., Network). The work in \citeSLR{4382127} provides a new technique for improving in-memory compression in order to minimize memory consumption and also a memory bank partitioning technique to cluster memory banks into active banks and inactive banks, which are powered off dynamically (i.e., Software). The work in \citeSLR{hosseini2012energy} propose two new energy-aware adaptation algorithms that reduce the cost by limiting light and transforming textures in video games (i.e., Software and Design).

The four studies address the same motivation (i.e., Battery life), but the first one focuses on \textit{Hardware}, the second focuses on \textit{Network}, the third one focuses on \textit{Software}, and the fourth one focuses on \textit{Software} and \textit{Design}. Similarly, the other identified motivations may be tackled using hardware, network, software, design, and/or art solutions.

\begin{figure}[t]
\centering
\begin{minipage}[b]{.45\textwidth}
\centering
\includegraphics[width=1\textwidth]{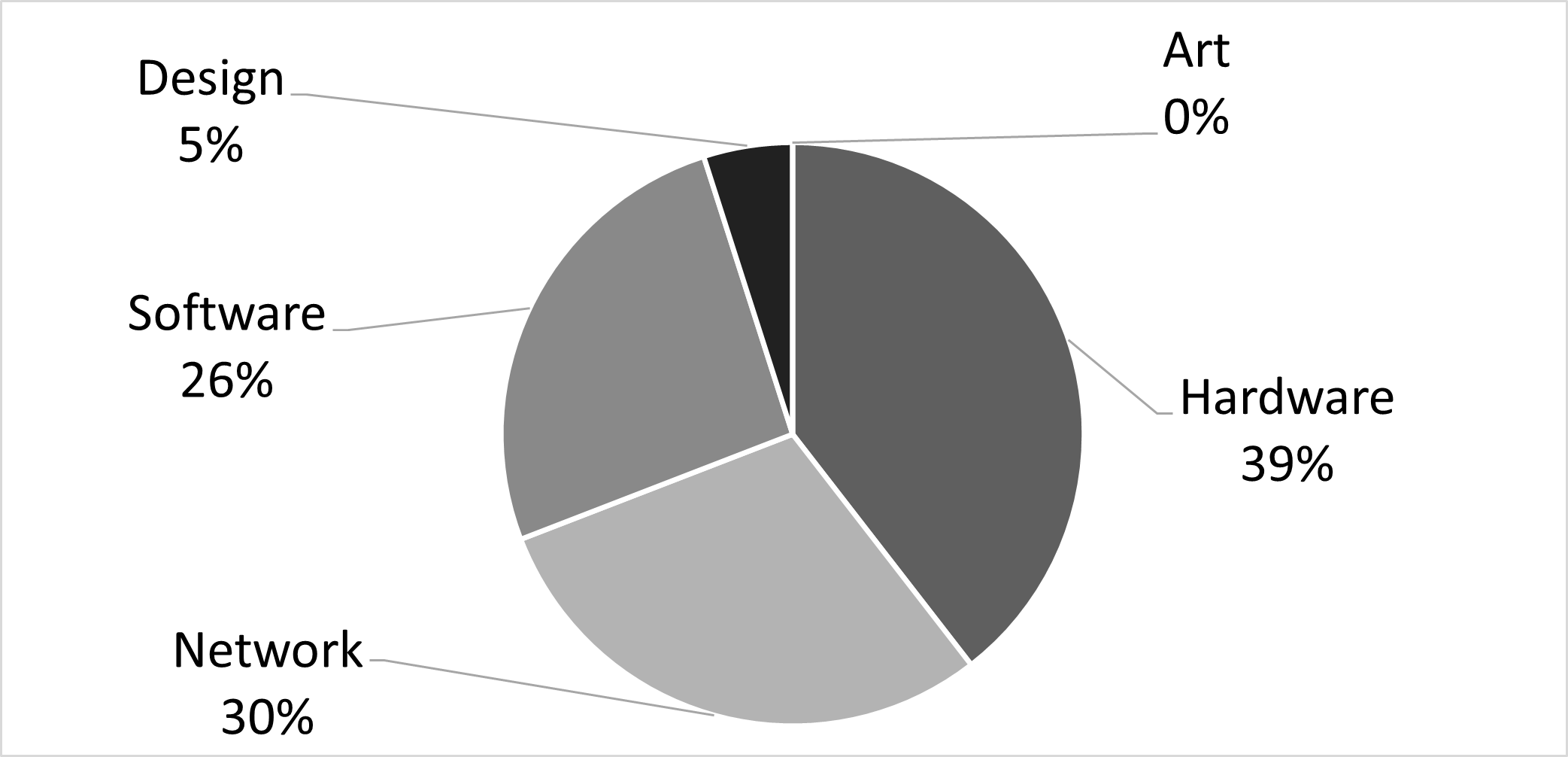}
\caption{Primary studies classified considering \\ layers in Green Video Games}
\label{Fig_RQ1_5}
\end{minipage}
\begin{minipage}[b]{.45\textwidth}
\centering
\includegraphics[width=1\textwidth]{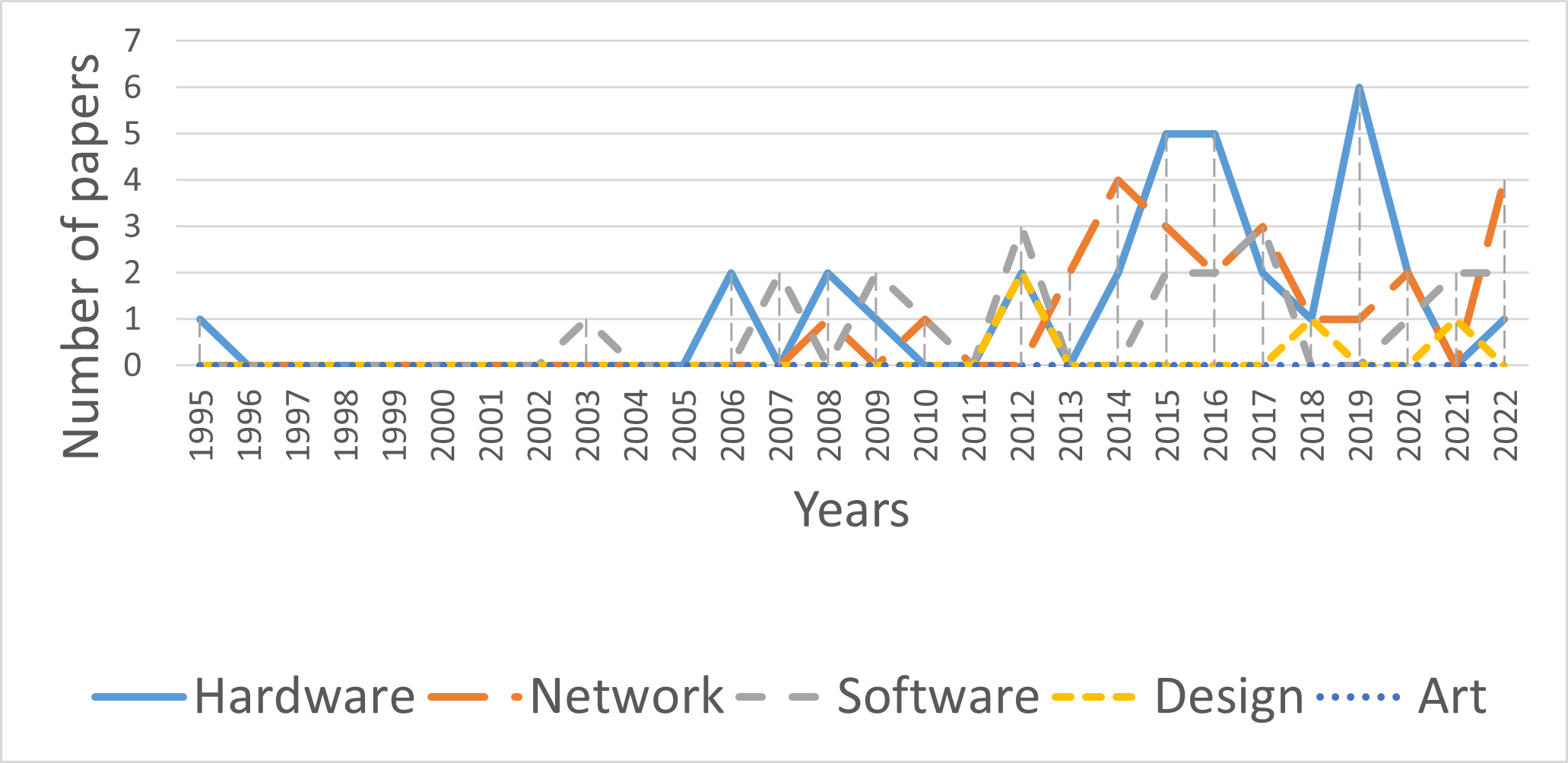}
\caption{Distribution of the layers in Green Video Games by years}
\label{Fig_RQ1_7}
\end{minipage}
\end{figure}

Fig. \ref{Fig_RQ1_5} shows the papers classified according to layers. Most of the primary studies focus on researching solutions through hardware components (39\% of the primary studies). Some primary studies focus on researching solutions through network and software components (30\% and 26\% of the primary studies, respectively). There are a few papers that focus on design components (5\% of the primary studies). However, there are no primary studies that focus on art components. 

Furthermore, Fig. \ref{Fig_RQ1_5} shows the number of papers in each layer by year. Note that there are no differences between the hardware, network, and software layers. The primary studies in these three layers emerged at approximately the same time. In the case of the design layer, studies have appeared sporadically, although it seems that their appearances have been more recurrent in recent years.

\begin{figure*}
\centering
\includegraphics[width=.98\textwidth]{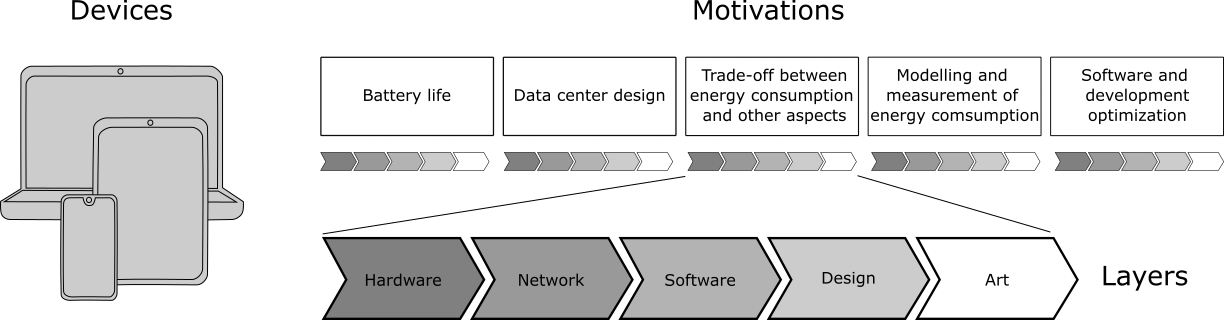}
\caption{Green Video Games issues regarding motivation, device, and layer}
\label{Fig_Our_Taxonomy}
\end{figure*}

Since the studies in Green Video Games are more motivation-oriented, they depend on the type of device and their solutions use components of specific layers. Therefore, we propose defining the issues around three axes: motivations, devices, and layers. Fig. \ref{Fig_Our_Taxonomy} summarizes the categories for each axe. The motivations and devices were identified from the primary studies and the layers were identified by professional video game developers in order to cover all of the relevant aspects of video game creation.

Table ~\ref{tab:our_taxonomy} shows an overview of the primary studies by issues. Most of the primary studies focus on Battery life (i.e., 36.23\% of the primary studies) or the Trade-off between energy consumption and other aspects (i.e., 36.23\% of the primary studies). In the case of Battery life, most of the studies provide solutions based on the hardware layer in mobile devices. In the case of the Trade-off between energy consumption and other aspects, most of the studies provide solutions based on the network layer in cloud computing and mobile devices. It is worth noting that these two issues are the only ones that contain some study in the design layer, even though all the studies are on mobile devices. 
 
\begin{table}[]
\caption{Overview of the issues in Green Video Games based on: motivation (1-Battery life, 2-Data center design, 3-Modelling and measurement of energy consumption, 4-General, 5-Software and development optimization, and 6-Trade-off between energy consumption and other aspects); layer (H-Hardware, N-Network, S-Software, D-Design, and A-Art); and device.}
\label{tab:our_taxonomy}
\setlength{\tabcolsep}{2.5pt}
\begin{tabular}{P{0.01\linewidth}l|m{0.12\linewidth}|m{0.08\linewidth}|m{0.03\linewidth}|m{0.04\linewidth}|m{0.04\linewidth}|m{0.19\linewidth}|m{0.08\linewidth}|m{0.04\linewidth}|m{0.04\linewidth}|m{0.04\linewidth}|m{0.04\linewidth}|m{0.12\linewidth}|}
\cline{3-14}
 & & \rot{Cloud computing} & \rot{Computers} & \rot{Edge devices} & \rot{Gaming devices} & \rot{Handheld devices} & \rot{Mobile devices} & \rot{Portable devices} & \rot{Reconfigurable MPSoCs} & \rot{Smart devices} & \rot{Wearable devices} & \rot{Wireless sensors} & \rot{Not specified} \\
 \hline 
\multicolumn{1}{|P{0.01\linewidth}}{\multirow{5}{=}{1}} & \multicolumn{1}{|l|}{H} & & & & & & \citeSLR{choi2021optimizing} \citeSLR{4749649} \citeSLR{7351775} \citeSLR{chen2015user} \citeSLR{peters2016frame} \citeSLR{pathania2014integrated} \citeSLR{dietrich2014lightweight} \citeSLR{pathania2015power} \citeSLR{park2017synergistic} & \citeSLR{Muhuri20182311} \citeSLR{gu2008control} \citeSLR{gu2006games} \citeSLR{gu2008power} & & & & & \\ \cline{2-14}
 \multicolumn{1}{|P{0.01\linewidth}}{} & \multicolumn{1}{|l|}{N} & \citeSLR{cai2013next} & & & & & \citeSLR{Ghergulescu2014} \citeSLR{zhang2017exploring} \citeSLR{8057058} \citeSLR{9816214} \citeSLR{cai2013next} & & & & & & \\\cline{2-14}
\multicolumn{1}{|P{0.01\linewidth}}{} & \multicolumn{1}{|l|}{S} & & & & & \citeSLR{4382127} & \citeSLR{4749649} \citeSLR{5679572} \citeSLR{thirugnanam2012dynamic} \citeSLR{pinto2017energy} \citeSLR{hosseini2012energy} \citeSLR{anand2009game} & & & & \citeSLR{pinto2017energy} & & \\\cline{2-14}
\multicolumn{1}{|P{0.01\linewidth}}{} & \multicolumn{1}{|l|}{D} & & & & & & \citeSLR{Yan201813} \citeSLR{hosseini2012energy} & & & & & & \\\cline{2-14}
\multicolumn{1}{|P{0.01\linewidth}}{} & \multicolumn{1}{|l|}{A} & & & & & & & & & & & & \\ \hline 

 \multicolumn{1}{|P{0.01\linewidth}}{\multirow{5}{=}{2}} & \multicolumn{1}{|l|}{H} & \citeSLR{Behiya202084} & & & & & & & & & & & \\\cline{2-14}
 \multicolumn{1}{|P{0.01\linewidth}}{} & \multicolumn{1}{|l|}{N} & \citeSLR{Behiya202084} \citeSLR{Dhib20222119} \citeSLR{nae2008efficient} & & & & & & & & & & & \\\cline{2-14}
 \multicolumn{1}{|P{0.01\linewidth}}{} & \multicolumn{1}{|l|}{S} & & & & & & & & & & & & \\\cline{2-14}
 \multicolumn{1}{|P{0.01\linewidth}}{} & \multicolumn{1}{|l|}{D} & & & & & & & & & & & & \\\cline{2-14}
 \multicolumn{1}{|P{0.01\linewidth}}{} & \multicolumn{1}{|l|}{A} & & & & & & & & & & & & \\ \hline 

 \multicolumn{1}{|P{0.01\linewidth}}{\multirow{5}{=}{3}} & \multicolumn{1}{|l|}{H} & & & & & & \citeSLR{chen2018user} \citeSLR{dietich2017estimating} \citeSLR{park2016hicap} & \citeSLR{6471008} & & \citeSLR{Song2019} & \citeSLR{Kone2016561} & &\\\cline{2-14}
\multicolumn{1}{|P{0.01\linewidth}}{} & \multicolumn{1}{|l|}{N} & \citeSLR{han2020virtual} & & & & \citeSLR{ma2017rofi} & \citeSLR{ma2017rofi} & & & & \citeSLR{Kone2016561} & & \\\cline{2-14}
\multicolumn{1}{|P{0.01\linewidth}}{} & \multicolumn{1}{|l|}{S} & & \citeSLR{Pereira2021} & & & & \citeSLR{park2016hicap} & & & & & & \\\cline{2-14}
\multicolumn{1}{|P{0.01\linewidth}}{} & \multicolumn{1}{|l|}{D} & & & & & & & & & & & & \\\cline{2-14}
\multicolumn{1}{|P{0.01\linewidth}}{} & \multicolumn{1}{|l|}{A} & & & & & & & & & & & & \\ \hline 

\multicolumn{1}{|P{0.01\linewidth}}{\multirow{5}{=}{4}} & \multicolumn{1}{|l|}{H} & & & & & & & & & & & & \\\cline{2-14}
\multicolumn{1}{|P{0.01\linewidth}}{} & \multicolumn{1}{|l|}{N} & \citeSLR{chuah2014cloud} & \citeSLR{8566409} & & & & & & & & & & \\\cline{2-14}
\multicolumn{1}{|P{0.01\linewidth}}{} & \multicolumn{1}{|l|}{S} & & \citeSLR{kasurinen2017concerns} & & & & \citeSLR{kasurinen2017concerns} & & & & & &\\\cline{2-14}
\multicolumn{1}{|P{0.01\linewidth}}{} & \multicolumn{1}{|l|}{D} & & & & & & & & & & & & \\\cline{2-14}
\multicolumn{1}{|P{0.01\linewidth}}{} & \multicolumn{1}{|l|}{A} & & & & & & & & & & & & \\ \hline 
\multicolumn{1}{|P{0.01\linewidth}}{\multirow{5}{=}{5}} & \multicolumn{1}{|l|}{H} & & & & & & & & & & & & \\\cline{2-14}
\multicolumn{1}{|P{0.01\linewidth}}{} & \multicolumn{1}{|l|}{N} & & & & & & & & & & & & \\\cline{2-14}
\multicolumn{1}{|P{0.01\linewidth}}{} & \multicolumn{1}{|l|}{S} & & \citeSLR{7380525} \citeSLR{1193041} & & & & & & & & & \citeSLR{LeBorgne201263} & \citeSLR{Rodriguez202218704} \\\cline{2-14}
\multicolumn{1}{|P{0.01\linewidth}}{} & \multicolumn{1}{|l|}{D} & & & & & & & & & & & & \\\cline{2-14}
\multicolumn{1}{|P{0.01\linewidth}}{} & \multicolumn{1}{|l|}{A} & & & & & & & & & & & & \\ \hline 

\multicolumn{1}{|P{0.01\linewidth}}{\multirow{5}{=}{6}} & \multicolumn{1}{|l|}{H} & \citeSLR{Guan20152434} & \citeSLR{Hertz2016142} & & \citeSLR{Nery201997} & \citeSLR{bose2012physics} & \citeSLR{Olivito201564} \citeSLR{Zhang20191546} \citeSLR{Teshima2015} \citeSLR{mochocki2006power} & &\citeSLR{Suriano2020136} & & & & \citeSLR{Cho2019499} \citeSLR{Toupas201988} \citeSLR{512397}\\ \cline{2-14}
\multicolumn{1}{|P{0.01\linewidth}}{} & \multicolumn{1}{|l|}{N} & \citeSLR{Kavalionak2015301} \citeSLR{Shea201561} \citeSLR{bhuyan2022end} \citeSLR{7037666} \citeSLR{Guan20152434} \citeSLR{ahmadi2014game} & & \citeSLR{bhuyan2022end} & & & \citeSLR{Jiang2022} \citeSLR{Zhang20191546} \citeSLR{lee2016exploiting} \citeSLR{6490354} \citeSLR{8566409} & & & & & & \\\cline{2-14}
\multicolumn{1}{|P{0.01\linewidth}}{} & \multicolumn{1}{|l|}{S} & \citeSLR{Gharsallaoui20171072} & \citeSLR{Hertz2016142} & & & & \citeSLR{Olivito201564} \citeSLR{Corbalan-Navarro20224375} \citeSLR{9591818} & & \citeSLR{Suriano2020136} & & & & \citeSLR{boeing2007evaluation} \\\cline{2-14}
\multicolumn{1}{|P{0.01\linewidth}}{} & \multicolumn{1}{|l|}{D} & & & & & & \citeSLR{9591818} & & & & & & \\\cline{2-14}
\multicolumn{1}{|P{0.01\linewidth}}{} & \multicolumn{1}{|l|}{A} & & & & & & & & & & & & \\ \hline 

\end{tabular}
\end{table}

The modelling and measurement of energy consumption is the next issue with more studies (i.e., 13.04\% of the primary studies). The studies tackling this issue focus on the hardware layer and mobile devices overall. The Software and deployment optimization is tackled by 5.80\% of the primary studies. Its studies focus on the software layer, but contrary to the other motivations, its studies focus on computers instead of mobile devices. Data center design is the issue with the fewest studies (i.e., 4.3\% of primary studies), and all of these studies tackle the issue through the network layer and cloud computing.

None of the issues are addressed through the art layer, and three studies address some issues through the design layers. This outcome will be taken into account for discussion. \textit{Should we discard these layers from the taxonomy?} \textit{Are these layers interesting for Green Video Games?} 

 
\subsection{RQ2. Which green techniques are applied for video game development?}

This question sheds light on green techniques that are applied in the context of video games. From the extracted data, we identified the following five techniques: 

\begin{itemize}
 \item \textbf{Cloud resource management} comprises the provisioning, allocation, and monitoring of the cloud resources (e.g., servers, memory, storage, network, CPU, application servers, virtual machines, etc) \cite{sobers2021introduction}. 
 \item \textbf{Content energy-aware adaptation} are techniques used to modify multimedia content in terms of reception and presentation in order to fit content to devices and at the same time minimize energy consumption \cite{ismail2013survey}.
 \item \textbf{CPU/GPU management} refers to techniques that take an integrated approach to study the influence of CPU and GPU on power consumption and performance in order to achieve better energy efficiency, for instance through synergistic frequency limiting policies \citeSLR{park2017synergistic}, or adapting the governing policies of both components taking into account user demands \citeSLR{chen2015user}.
 \item \textbf{Dynamic voltage and frequency scaling (DVFS)} is a technique that aims at reducing the dynamic power consumption by dynamically adjusting the voltage and frequency of a CPU \cite{jha2001low,cardoso2017embedded}. 
 \item \textbf{Field-programmable gate arrays (FPGAs)} are integrated circuits that can be configured and reprogrammed by the customer. Some platforms are developed using FPGAs because they deliver higher energy efficiency than CPU-only or GPU-based platforms \citeSLR{Cho2019499}.
\end{itemize}

\begin{figure*}
\centering
\includegraphics[width=.45\textwidth]{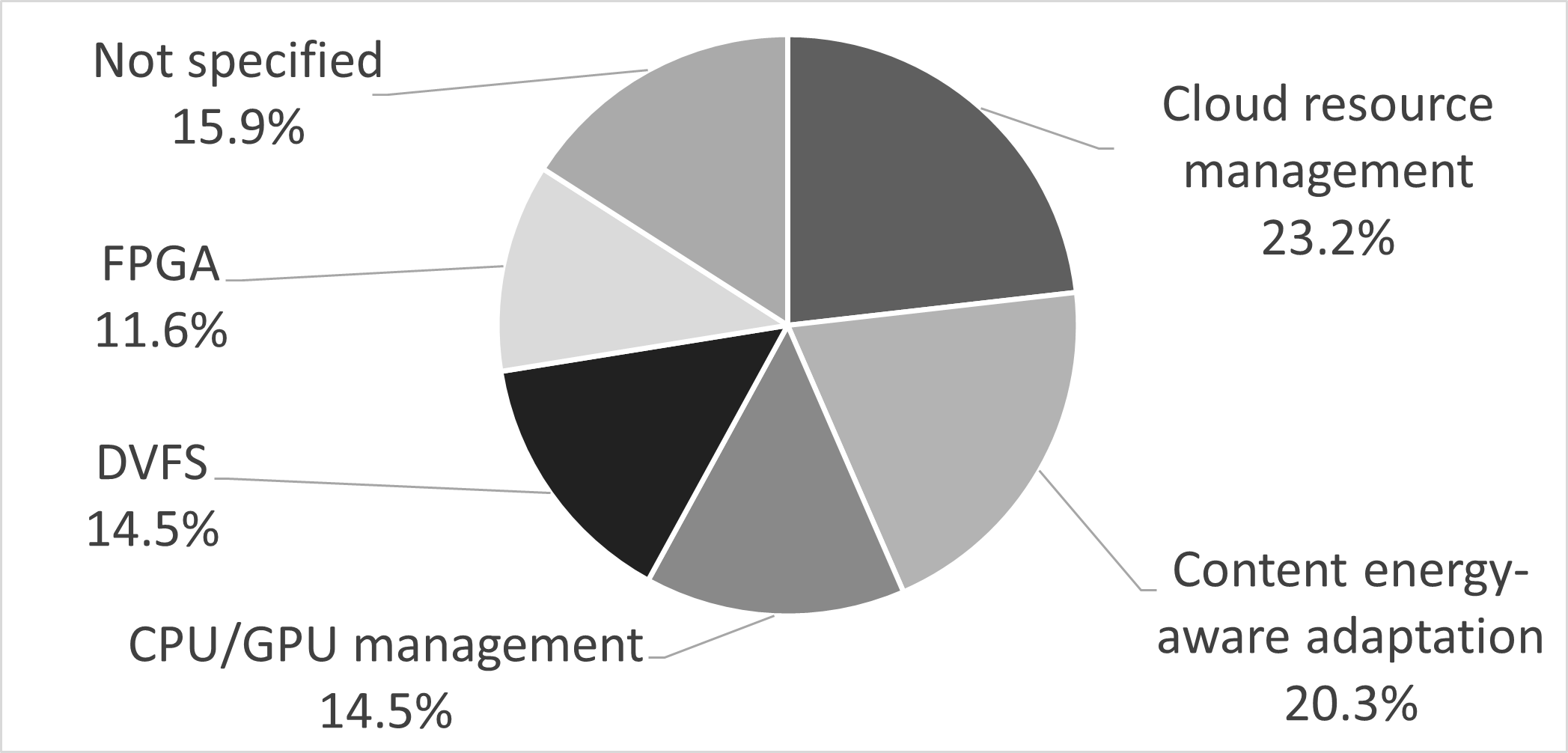}
\caption{Distribution of the primary studies based on technique}
\label{Fig_RQ2_1}
\end{figure*}

Fig. \ref{Fig_RQ2_1} shows the number of primary studies for each technique. The most common techniques are Cloud resource management and Content energy-aware adaptation. Nevertheless, the difference between the number of papers for each technique is not large. Perhaps the most relevant aspect of this figure is the number of primary studies that do not specify the technique applied.

In addition, to answer RQ2, we studied these techniques based on the issues identified in RQ1. In terms of the motivation (Table \ref{tab:RQ2_1}), it can be observed that the five techniques are used for both Battery life and the Trade-off between energy consumption and other aspects. For Modelling and measurement of energy consumption, researchers are using four of the five techniques (i.e., Cloud resource management, Content energy-aware adaptation, CPU/GPU management, and DVFS). For Software and development optimization, researchers are using two of the five techniques (i.e., Content-energy aware adaptation and CPU/GPU management). However, for Data center design, the only technique being applied is Cloud resource management.

\begin{table}[]
\centering
\renewcommand{\arraystretch}{1.2}
\caption{Techniques used in primary studies over motivation (1-Battery life, 2-Data center design, 3-Modelling and measurement of energy consumption, 4-General, 5-Software and development optimization, and 6-Trade-off between energy consumption and other aspects).}
\label{tab:RQ2_1}
\begin{tabular}{p{0.01\linewidth}|p{0.16\linewidth}|p{0.16\linewidth}|p{0.12\linewidth}|p{0.12\linewidth}|p{0.12\linewidth}|p{0.12\linewidth}|}
\cline{2-7}
 & Cloud resource management & Content energy-aware adaptation & CPU/GPU management & DVFS & FPGA & Not specified \\ \hline
\multicolumn{1}{|p{0.01\linewidth}|}{1} & \citeSLR{cai2013next} & \citeSLR{Ghergulescu2014} \citeSLR{zhang2017exploring} \citeSLR{5679572} \citeSLR{8057058} \citeSLR{9816214} \citeSLR{thirugnanam2012dynamic} \citeSLR{hosseini2012energy} \citeSLR{anand2009game} & \citeSLR{7351775} \citeSLR{chen2015user} \citeSLR{pathania2014integrated} \citeSLR{gu2008power} \citeSLR{park2017synergistic} & \citeSLR{Muhuri20182311} \citeSLR{4749649} \citeSLR{gu2008control} \citeSLR{peters2016frame} \citeSLR{gu2006games} \citeSLR{dietrich2014lightweight} & \citeSLR{choi2021optimizing} \citeSLR{pathania2015power} & \citeSLR{Yan201813} \citeSLR{4382127} \citeSLR{pinto2017energy} \\\hline
\multicolumn{1}{|p{0.01\linewidth}|}{2} & \citeSLR{Behiya202084} \citeSLR{Dhib20222119} \citeSLR{nae2008efficient} & & & & & \\\hline
\multicolumn{1}{|p{0.01\linewidth}|}{3} & \citeSLR{Kone2016561} \citeSLR{han2020virtual} & \citeSLR{ma2017rofi} & \citeSLR{6471008} \citeSLR{park2016hicap} & \citeSLR{Song2019} \citeSLR{chen2018user} \citeSLR{dietich2017estimating} & & \citeSLR{Pereira2021} \\\hline
\multicolumn{1}{|p{0.01\linewidth}|}{4} & \citeSLR{chuah2014cloud} \citeSLR{8566409} & & & & & \citeSLR{kasurinen2017concerns} \\\hline
\multicolumn{1}{|p{0.01\linewidth}|}{5} & & \citeSLR{LeBorgne201263} \citeSLR{1193041} & \citeSLR{7380525} & & & \citeSLR{Rodriguez202218704} \\\hline
\multicolumn{1}{|p{0.01\linewidth}|}{6} & \citeSLR{Kavalionak2015301} \citeSLR{Shea201561} \citeSLR{Jiang2022} \citeSLR{bhuyan2022end} \citeSLR{6490354} \citeSLR{7037666} \citeSLR{Guan20152434} \citeSLR{8566409} & \citeSLR{Gharsallaoui20171072} \citeSLR{Corbalan-Navarro20224375} \citeSLR{ahmadi2014game} & \citeSLR{Suriano2020136} \citeSLR{lee2016exploiting} & \citeSLR{mochocki2006power} & \citeSLR{Olivito201564} \citeSLR{Zhang20191546} \citeSLR{Cho2019499} \citeSLR{Nery201997} \citeSLR{Toupas201988} \citeSLR{bose2012physics} & \citeSLR{Teshima2015} \citeSLR{512397} \citeSLR{Hertz2016142} \citeSLR{9591818} \citeSLR{boeing2007evaluation}
\\\hline
\end{tabular}
\end{table}

In terms of the device (Table \ref{tab:RQ2_2}), it can be observed that the five techniques are used for Mobile devices. In contrast, for Cloud computing, the technique that is most used is Cloud resource management. Also noteworthy is the number of studies that did not specify any technique. It is striking that most of the papers that did not specify the device also did not specify the technique.

\begin{table}[]
\centering
\renewcommand{\arraystretch}{1.2}
\caption{Techniques used in primary studies over devices (D1-Cloud computing, D2-Computers, D3-Edge devices, D4-Gaming devices, D5-Handheld devices, D6-Mobile devices, D7-Portable devices, D8-Reconfigurable MPSoCs, D9-Smart devices, D10-Wearable devices, D11-Wireless sensors, and D12-Not specified).}
\label{tab:RQ2_2}
\begin{tabular}{p{0.03\linewidth}|p{0.16\linewidth}|p{0.16\linewidth}|p{0.12\linewidth}|p{0.12\linewidth}|p{0.12\linewidth}|p{0.12\linewidth}|}
\cline{2-7}
 & Cloud resource management & Content energy-aware adaptation & CPU/GPU management & DVFS & FPGA & Not specified \\ \hline
\multicolumn{1}{|p{0.03\linewidth}|}{D1} & \citeSLR{Kavalionak2015301} \citeSLR{Behiya202084} \citeSLR{Shea201561} \citeSLR{chuah2014cloud} \citeSLR{bhuyan2022end} \citeSLR{7037666} \citeSLR{Dhib20222119} \citeSLR{Guan20152434} \citeSLR{nae2008efficient} \citeSLR{cai2013next} \citeSLR{han2020virtual} & \citeSLR{Gharsallaoui20171072} \citeSLR{ahmadi2014game} & & & & \\\hline
\multicolumn{1}{|p{0.03\linewidth}|}{D2} & \citeSLR{8566409} & \citeSLR{1193041} & \citeSLR{7380525} & & & \citeSLR{Pereira2021} \citeSLR{kasurinen2017concerns} \citeSLR{Hertz2016142} \\\hline
\multicolumn{1}{|p{0.03\linewidth}|}{D3} & \citeSLR{bhuyan2022end} & & & & & \\\hline
\multicolumn{1}{|p{0.03\linewidth}|}{D4} & & & & & \citeSLR{Nery201997} & \\\hline
\multicolumn{1}{|p{0.03\linewidth}|}{D5} & & \citeSLR{ma2017rofi} & & &\citeSLR{bose2012physics} & \citeSLR{4382127} \\\hline
\multicolumn{1}{|p{0.03\linewidth}|}{D6} & \citeSLR{Jiang2022} \citeSLR{6490354} \citeSLR{8566409} \citeSLR{cai2013next} & \citeSLR{Ghergulescu2014} \citeSLR{ma2017rofi} \citeSLR{zhang2017exploring} \citeSLR{5679572} \citeSLR{8057058} \citeSLR{Corbalan-Navarro20224375} \citeSLR{9816214} \citeSLR{thirugnanam2012dynamic} \citeSLR{hosseini2012energy} \citeSLR{anand2009game} & \citeSLR{lee2016exploiting} \citeSLR{7351775} \citeSLR{chen2015user} \citeSLR{park2016hicap} \citeSLR{pathania2014integrated} \citeSLR{park2017synergistic} & \citeSLR{chen2018user} \citeSLR{4749649} \citeSLR{dietich2017estimating} \citeSLR{peters2016frame} \citeSLR{dietrich2014lightweight} \citeSLR{mochocki2006power} & \citeSLR{Olivito201564} \citeSLR{Zhang20191546} \citeSLR{choi2021optimizing} \citeSLR{pathania2015power} & \citeSLR{Teshima2015} \citeSLR{Yan201813} \citeSLR{kasurinen2017concerns} \citeSLR{9591818} \citeSLR{pinto2017energy} \\\hline
\multicolumn{1}{|p{0.03\linewidth}|}{D7} & & & \citeSLR{6471008} \citeSLR{gu2008power} & \citeSLR{Muhuri20182311} \citeSLR{gu2008control} \citeSLR{gu2006games} & & \\\hline
\multicolumn{1}{|p{0.03\linewidth}|}{D8} & & & \citeSLR{Suriano2020136} & & & \\\hline
\multicolumn{1}{|p{0.03\linewidth}|}{D9} & & & & \citeSLR{Song2019} & & \\\hline
\multicolumn{1}{|p{0.03\linewidth}|}{D10} & \citeSLR{Kone2016561} & & & & & \citeSLR{pinto2017energy} \\\hline
\multicolumn{1}{|p{0.03\linewidth}|}{D11} & & \citeSLR{LeBorgne201263} & & & & \\\hline
\multicolumn{1}{|p{0.03\linewidth}|}{D12} & & & & & \citeSLR{Cho2019499} \citeSLR{Toupas201988} & \citeSLR{Rodriguez202218704} \citeSLR{512397} \citeSLR{boeing2007evaluation} \\\hline
\end{tabular}
\end{table}

In terms of the layer (Table \ref{tab:RQ2_3}), it can be observed that most techniques are not only used in one specific layer. However, there are techniques that are more commonly used on one layer, such as Cloud resource management. Cloud Resource management is more frequently used in solutions on the Network layer. 

In contrast, DVFS and FPGA are used exclusively in the Hardware layer. In Table \ref{tab:RQ2_3}, one study that uses DVFS is classified in the Software layer. Similarly, two studies that use FPGA are classified in the Network layer and the Software layer. However, we can also find these studies in the Hardware layer. These three studies provide solutions where part of this solution is based on Hardware and part of this solution is based on Network or Software. DVFS and FPGA are always used in the solution part that is based on Hardware. Therefore, although these studies appear in Network and Software, DVFS and FPGA are always used in the Hardware layer.

\begin{table}[]
\centering
\renewcommand{\arraystretch}{1.2}
\caption{Techniques used in primary studies over layers (H-Hardware, N-Network, S-Software, D-Design, and A-Art).}
\label{tab:RQ2_3}
\begin{tabular}{p{0.02\linewidth}|p{0.16\linewidth}|p{0.13\linewidth}|p{0.12\linewidth}|p{0.16\linewidth}|p{0.12\linewidth}|p{0.12\linewidth}|}
\cline{2-7}
 & Cloud resource management & Content energy-aware adaptation & CPU/GPU management & DVFS & FPGA & Not specified \\ \hline
\multicolumn{1}{|p{0.02\linewidth}|}{H} & \citeSLR{Behiya202084} \citeSLR{Kone2016561} \citeSLR{Guan20152434} & & \citeSLR{Suriano2020136} \citeSLR{6471008} \citeSLR{7351775} \citeSLR{chen2015user} \citeSLR{park2016hicap} \citeSLR{pathania2014integrated} \citeSLR{gu2008power} \citeSLR{park2017synergistic} & \citeSLR{Muhuri20182311} \citeSLR{Song2019} \citeSLR{chen2018user} \citeSLR{4749649} \citeSLR{gu2008control} \citeSLR{dietich2017estimating} \citeSLR{peters2016frame} \citeSLR{gu2006games} \citeSLR{dietrich2014lightweight} \citeSLR{mochocki2006power} & \citeSLR{Olivito201564} \citeSLR{Zhang20191546} \citeSLR{Cho2019499} \citeSLR{Nery201997} \citeSLR{Toupas201988} \citeSLR{choi2021optimizing} \citeSLR{bose2012physics} \citeSLR{pathania2015power} & \citeSLR{Teshima2015} \citeSLR{512397} \citeSLR{Hertz2016142} \\\hline
\multicolumn{1}{|p{0.02\linewidth}|}{N} & \citeSLR{Kavalionak2015301} \citeSLR{Behiya202084} \citeSLR{Shea201561} \citeSLR{Jiang2022} \citeSLR{Kone2016561} \citeSLR{chuah2014cloud} \citeSLR{bhuyan2022end} \citeSLR{8566409} \citeSLR{6490354} \citeSLR{7037666} \citeSLR{Dhib20222119} \citeSLR{Guan20152434} \citeSLR{8566409} \citeSLR{nae2008efficient} \citeSLR{cai2013next} \citeSLR{han2020virtual} & \citeSLR{Ghergulescu2014} \citeSLR{ma2017rofi} \citeSLR{zhang2017exploring} \citeSLR{8057058} \citeSLR{9816214} \citeSLR{ahmadi2014game} & \citeSLR{lee2016exploiting} & & \citeSLR{Zhang20191546} & \\\hline
\multicolumn{1}{|p{0.02\linewidth}|}{S} & & \citeSLR{LeBorgne201263} \citeSLR{Gharsallaoui20171072} \citeSLR{1193041} \citeSLR{5679572} \citeSLR{Corbalan-Navarro20224375} \citeSLR{thirugnanam2012dynamic} \citeSLR{hosseini2012energy} \citeSLR{anand2009game} & \citeSLR{Suriano2020136} \citeSLR{7380525} \citeSLR{park2016hicap} & \citeSLR{4749649} & \citeSLR{Olivito201564} & \citeSLR{Pereira2021} \citeSLR{Rodriguez202218704} \citeSLR{kasurinen2017concerns} \citeSLR{4382127} \citeSLR{Hertz2016142} \citeSLR{9591818} \citeSLR{pinto2017energy} \citeSLR{boeing2007evaluation} \\\hline
\multicolumn{1}{|p{0.02\linewidth}|}{D} & & \citeSLR{hosseini2012energy} & \citeSLR{6471008} & & & \citeSLR{Yan201813} \citeSLR{9591818} \\\hline
\multicolumn{1}{|p{0.02\linewidth}|}{A} & & & & & & \\\hline
\end{tabular}
\end{table}

In the literature, there are some research works that focus on categorizing the Green Computing techniques. The work in \cite{choudhary2014survey} presents a classification for the Green Computing techniques based on four different levels: the hardware and firmware level, the operating system level, the virtualization level, and the data center level. The work in \cite{shuja2017greening} presents a review of Green Computing techniques amidst the emerging computing technologies that are evident in our society. However, these works only provide descriptions of what type of techniques are employed for the different levels or categories, and they do not go into what specific techniques exist. For this reason, it is difficult to fairly compare if the techniques used in Green Computing are the same ones used in video games. 
 

\subsection{RQ3. What are the limitations of the current Green Video Game studies?}

In this section, we outline the limitations of the existing studies in Green Video Games. We also studied the evidence level, which is used to check the maturity of the technique, practice, solution, or method presented in the paper. 

None of the studies provide threats to validity. Threats to validity would be those internal and external factors that may prevent one from measuring what one wants to measure or that may obscure the relationship between the dependent and the independent variables. In other words, threats to validity help to replicate the experiment or evaluation using the same conditions in order to compare the results with other studies fairly. However, despite the importance of the threats to validity, 100\% of the primary studies have not included them.

Since there were no threats to validity, we focused on extracting any limitation, not only the ones related to the validation. Only four studies describe some limitation of their work. The work in \citeSLR{1193041} points out two limitations of the proposed method: 1) the probability of error can be high if the environment contains a lot of small objects, and 2) the application of the method to scenes with rigid objects only is limited. These limitations are shared with other postfiltering methods. In \citeSLR{anand2009game}, the authors only consider the wireless network interface card component of the Personal Digital Assistant (PDA). In \citeSLR{pathania2014integrated}, since the authors were not aware of any available record and replay mechanism for 3D games on the Android platform, they requested volunteers to play the main level of the games with repetition of their game-play as far as possible. The work in \citeSLR{dietrich2014lightweight} uses the processor of a Samsung Galaxy Nexus for the evaluation, but this processor is configured such that the CPU performance counters cannot be read. To overcome this limitation, the authors ran the games at all available processing frequencies and measured the frame rate as well as the time spent in the loading state.

The extracted data also shows that the following topics are recurrently reported as future works:

\begin{itemize}
\item Generalization of the results evaluating the solution on other devices \citeSLR{Muhuri20182311} \citeSLR{Zhang20191546} \citeSLR{park2016hicap} \citeSLR{5679572}.
\item Additional parameters or parameter tuning \citeSLR{zhang2017exploring} \citeSLR{6490354} \citeSLR{8057058} \citeSLR{gu2008control} \citeSLR{han2020virtual}.
\item Testing the solution with other games or different video game genres \citeSLR{thirugnanam2012dynamic} \citeSLR{7380525} \citeSLR{nae2008efficient}.
\item Extending the solution using machine-learning techniques \citeSLR{park2016hicap} \citeSLR{park2017synergistic}.
\item Integrating or improving solutions in the cloud \citeSLR{Kavalionak2015301} \citeSLR{Gharsallaoui20171072} \citeSLR{ahmadi2014game}.
\end{itemize}

\begin{figure*}
\centering
\includegraphics[width=.45\textwidth]{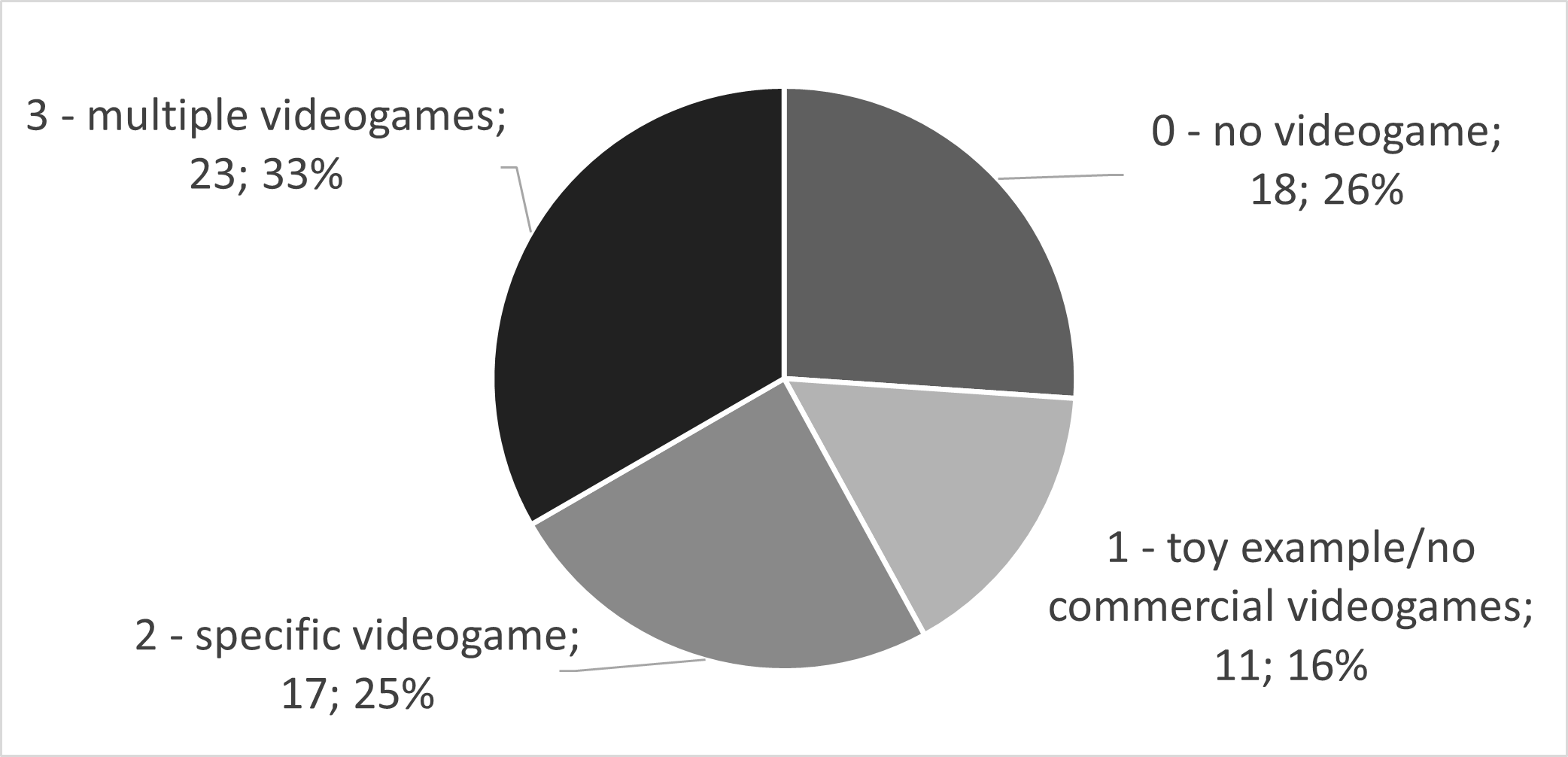}
\caption{Distribution of the primary studies based on evidence level}
\label{Fig_RQ3_1}
\end{figure*}

In addition to the limitations, this section also provides information about the evidence level. Fig. \ref{Fig_RQ3_1} shows the distribution of evidence levels described in Section \ref{sec:data_collection}. This figure shows both the number of papers and the percentage of papers for each evidence level. Taking into account this figure, most of the studies have been evaluated using at least one specific video game. 

In terms of the motivation (Table \ref{tab:RQ3_1}, most primary studies have been tested with at least one specific video game, or even with several video games. The exceptions, however, are the studies focusing on the Trade-off between energy consumption and other aspects and the studies offering General perspectives. In these two cases, the number of studies evaluated on toy examples, no-commercial video games, or that do not use video games exceeds that of studies with evaluations based on commercial games.

\begin{table}[]
\centering
\renewcommand{\arraystretch}{1.2}
\caption{Level of evidence of the primary studies over motivation (1-Battery life, 2-Data center design, 3-Modelling and measurement of energy consumption, 4-General, 5-Software and development optimization, and 6-Trade-off between energy consumption and other aspects).}
\label{tab:RQ3_1}
\begin{tabular}{p{0.01\linewidth}|p{0.20\linewidth}|p{0.23\linewidth}|p{0.20\linewidth}|p{0.24\linewidth}|}
 \cline{2-5}
 & 0 - no video game & 1 - toy example/no commercial video games & 2 - specific video game & 3 - multiple video games \\ \hline
\multicolumn{1}{|p{0.01\linewidth}|}{1} & \citeSLR{4382127} \citeSLR{4749649} \citeSLR{9816214} \citeSLR{pinto2017energy} \citeSLR{cai2013next} & \citeSLR{Yan201813} & \citeSLR{Ghergulescu2014} \citeSLR{5679572} \citeSLR{gu2008control} \citeSLR{thirugnanam2012dynamic} \citeSLR{anand2009game} \citeSLR{gu2006games} \citeSLR{gu2008power} & \citeSLR{Muhuri20182311} \citeSLR{choi2021optimizing} \citeSLR{zhang2017exploring} \citeSLR{7351775} \citeSLR{8057058} \citeSLR{chen2015user} \citeSLR{hosseini2012energy} \citeSLR{peters2016frame} \citeSLR{pathania2014integrated} \citeSLR{dietrich2014lightweight} \citeSLR{pathania2015power} \citeSLR{park2017synergistic} \\ \hline
\multicolumn{1}{|p{0.01\linewidth}|}{2} & \citeSLR{Behiya202084} & & \citeSLR{Dhib20222119} \citeSLR{nae2008efficient} & \\ \hline
\multicolumn{1}{|p{0.01\linewidth}|}{3} & \citeSLR{Kone2016561} \citeSLR{Song2019} \citeSLR{6471008} & \citeSLR{Pereira2021} & & \citeSLR{chen2018user} \citeSLR{ma2017rofi} \citeSLR{dietich2017estimating} \citeSLR{park2016hicap} \citeSLR{han2020virtual} \\ \hline
\multicolumn{1}{|p{0.01\linewidth}|}{4} & \citeSLR{8566409} & \citeSLR{chuah2014cloud} & & \citeSLR{kasurinen2017concerns} \\ \hline
\multicolumn{1}{|p{0.01\linewidth}|}{5} & \citeSLR{Rodriguez202218704} & \citeSLR{1193041} & \citeSLR{LeBorgne201263} \citeSLR{7380525} & \\ \hline
\multicolumn{1}{|p{0.01\linewidth}|}{6}& \citeSLR{Jiang2022} \citeSLR{Teshima2015} \citeSLR{Toupas201988} \citeSLR{6490354} \citeSLR{512397} \citeSLR{Hertz2016142} \citeSLR{boeing2007evaluation} & \citeSLR{Kavalionak2015301} \citeSLR{Zhang20191546} \citeSLR{lee2016exploiting} \citeSLR{bhuyan2022end} \citeSLR{9591818} \citeSLR{bose2012physics} \citeSLR{mochocki2006power} & \citeSLR{Suriano2020136} \citeSLR{Shea201561} \citeSLR{Olivito201564} \citeSLR{Gharsallaoui20171072} \citeSLR{7037666} \citeSLR{8566409} & \citeSLR{Cho2019499} \citeSLR{Nery201997} \citeSLR{Corbalan-Navarro20224375} \citeSLR{Guan20152434} \citeSLR{ahmadi2014game} \\  \hline
\end{tabular}
\end{table}

In terms of the device (Table \ref{tab:RQ3_2}), the number of studies evaluated on video games stands out, especially in the case of mobile devices. In the case of cloud computing and portable devices, most studies were evaluated on a single commercial video game. However, most of the studies, where the device is not specified, conducted their evaluations without using a video game.

\begin{table}[]
\centering
\renewcommand{\arraystretch}{1.2}
\caption{Level of evidence of the primary studies over devices (D1-Cloud computing, D2-Computers, D3-Edge devices, D4-Gaming devices, D5-Handheld devices, D6-Mobile devices, D7-Portable devices, D8-Reconfigurable MPSoCs, D9-Smart devices, D10-Wearable devices, D11-Wireless sensors, and D12-Not specified).}
\label{tab:RQ3_2}
\begin{tabular}{p{0.03\linewidth}|p{0.20\linewidth}|p{0.23\linewidth}|p{0.20\linewidth}|p{0.24\linewidth}|}
 \cline{2-5}
 & 0 - no video game & 1 - toy example/no commercial video games & 2 - specific video game & 3 - multiple video games \\ \hline
\multicolumn{1}{|p{0.03\linewidth}|}{D1} & \citeSLR{Behiya202084} \citeSLR{cai2013next} & \citeSLR{Kavalionak2015301} \citeSLR{chuah2014cloud} \citeSLR{bhuyan2022end} & \citeSLR{Shea201561} \citeSLR{Gharsallaoui20171072} \citeSLR{7037666} \citeSLR{Dhib20222119} \citeSLR{nae2008efficient} & \citeSLR{Guan20152434} \citeSLR{ahmadi2014game} \citeSLR{han2020virtual} \\ \hline
\multicolumn{1}{|p{0.03\linewidth}|}{D2} & \citeSLR{8566409} \citeSLR{Hertz2016142} & \citeSLR{Pereira2021} \citeSLR{1193041} & \citeSLR{7380525} & \citeSLR{kasurinen2017concerns} \\ \hline
\multicolumn{1}{|p{0.03\linewidth}|}{D3} & & \citeSLR{bhuyan2022end} & & \\ \hline
\multicolumn{1}{|p{0.03\linewidth}|}{D4} & & & & \citeSLR{Nery201997} \\ \hline
\multicolumn{1}{|p{0.03\linewidth}|}{D5} & \citeSLR{4382127} & \citeSLR{bose2012physics} & & \citeSLR{ma2017rofi} \\ \hline
\multicolumn{1}{|p{0.03\linewidth}|}{D6} & \citeSLR{Jiang2022} \citeSLR{Teshima2015} \citeSLR{6490354} \citeSLR{4749649} \citeSLR{9816214} \citeSLR{pinto2017energy} \citeSLR{cai2013next} & \citeSLR{Zhang20191546} \citeSLR{Yan201813} \citeSLR{lee2016exploiting} \citeSLR{9591818} \citeSLR{mochocki2006power} & \citeSLR{Olivito201564} \citeSLR{Ghergulescu2014} \citeSLR{5679572} \citeSLR{8566409} \citeSLR{thirugnanam2012dynamic} \citeSLR{anand2009game} & \citeSLR{kasurinen2017concerns} \citeSLR{choi2021optimizing} \citeSLR{chen2018user} \citeSLR{ma2017rofi} \citeSLR{zhang2017exploring} \citeSLR{7351775} \citeSLR{8057058} \citeSLR{Corbalan-Navarro20224375} \citeSLR{chen2015user} \citeSLR{hosseini2012energy} \citeSLR{dietich2017estimating} \citeSLR{peters2016frame} \citeSLR{park2016hicap} \citeSLR{pathania2014integrated} \citeSLR{dietrich2014lightweight} \citeSLR{pathania2015power} \citeSLR{park2017synergistic} \\ \hline
\multicolumn{1}{|p{0.03\linewidth}|}{D7} & \citeSLR{6471008} & & \citeSLR{gu2008control} \citeSLR{gu2006games} \citeSLR{gu2008power} & \citeSLR{Muhuri20182311} \\ \hline
\multicolumn{1}{|p{0.03\linewidth}|}{D8} & & & \citeSLR{Suriano2020136} & \\ \hline
\multicolumn{1}{|p{0.03\linewidth}|}{D9} & \citeSLR{Song2019} & & & \\ \hline
\multicolumn{1}{|p{0.03\linewidth}|}{D10} & \citeSLR{Kone2016561} \citeSLR{pinto2017energy} & & & \\ \hline
\multicolumn{1}{|p{0.03\linewidth}|}{D11} & & & \citeSLR{LeBorgne201263} & \\ \hline
\multicolumn{1}{|p{0.03\linewidth}|}{D12} & \citeSLR{Toupas201988} \citeSLR{Rodriguez202218704} \citeSLR{512397} \citeSLR{boeing2007evaluation} & & & \citeSLR{Cho2019499} \\  \hline
\end{tabular}
\end{table}

In terms of the layer (Table \ref{tab:RQ3_3}), there was a fairly similar number of studies at all levels of evidence, although the number of studies evaluated on toy examples or non-commercial video games was somewhat lower. The result that is a bit striking is the progression in the studies evaluated on multiple video games. As you move up the layer, there are fewer studies evaluated on multiple video games. In other words, the hardware layer has more studies evaluated with multiple games than the network layer; the network layer has more studies evaluated with multiple games than the software layer; the software layer has more studies evaluated with multiple games than the design layer; and the design layer has more studies evaluated with multiple games than the art layer.


\begin{table}[]
\centering
\renewcommand{\arraystretch}{1.2}
\caption{Level of evidence of the primary studies over layers (H-Hardware, N-Network, S-Software, D-Design, and A-Art).}
\label{tab:RQ3_3}
\begin{tabular}{p{0.02\linewidth}|p{0.20\linewidth}|p{0.23\linewidth}|p{0.20\linewidth}|p{0.24\linewidth}|}
 \cline{2-5}
 & 0 - no video game & 1 - toy example/no commercial video 
games & 2 - specific video game & 3 - multiple video games \\ \hline
\multicolumn{1}{|p{0.02\linewidth}|}{H} & \citeSLR{Behiya202084} \citeSLR{Kone2016561} \citeSLR{Song2019} \citeSLR{Teshima2015} \citeSLR{Toupas201988} \citeSLR{4749649} \citeSLR{512397} \citeSLR{6471008} \citeSLR{Hertz2016142} & \citeSLR{Zhang20191546} \citeSLR{bose2012physics} \citeSLR{mochocki2006power} & \citeSLR{Suriano2020136} \citeSLR{Olivito201564} \citeSLR{gu2008control} \citeSLR{gu2006games} \citeSLR{gu2008power} & \citeSLR{Muhuri20182311} \citeSLR{Cho2019499} \citeSLR{Nery201997} \citeSLR{choi2021optimizing} \citeSLR{chen2018user} \citeSLR{7351775} \citeSLR{Guan20152434} \citeSLR{chen2015user} \citeSLR{dietich2017estimating} \citeSLR{peters2016frame} \citeSLR{park2016hicap} \citeSLR{pathania2014integrated} \citeSLR{dietrich2014lightweight} \citeSLR{pathania2015power} \citeSLR{park2017synergistic} \\ \hline
\multicolumn{1}{|p{0.02\linewidth}|}{N} & \citeSLR{Behiya202084} \citeSLR{Jiang2022} \citeSLR{Kone2016561} \citeSLR{8566409} \citeSLR{6490354} \citeSLR{9816214} \citeSLR{cai2013next} & \citeSLR{Kavalionak2015301} \citeSLR{Zhang20191546} \citeSLR{chuah2014cloud} \citeSLR{lee2016exploiting} \citeSLR{bhuyan2022end} & \citeSLR{Shea201561} \citeSLR{Ghergulescu2014} \citeSLR{7037666} \citeSLR{Dhib20222119} \citeSLR{8566409} \citeSLR{nae2008efficient} & \citeSLR{ma2017rofi} \citeSLR{zhang2017exploring} \citeSLR{8057058} \citeSLR{Guan20152434} \citeSLR{ahmadi2014game} \citeSLR{han2020virtual} \\ \hline
\multicolumn{1}{|p{0.02\linewidth}|}{S}  & \citeSLR{Rodriguez202218704} \citeSLR{4382127} \citeSLR{4749649} \citeSLR{Hertz2016142} \citeSLR{pinto2017energy} \citeSLR{boeing2007evaluation} & \citeSLR{Pereira2021} \citeSLR{1193041} \citeSLR{9591818} & \citeSLR{Suriano2020136} \citeSLR{LeBorgne201263} \citeSLR{Olivito201564} \citeSLR{Gharsallaoui20171072} \citeSLR{7380525} \citeSLR{5679572} \citeSLR{thirugnanam2012dynamic} \citeSLR{anand2009game} & \citeSLR{kasurinen2017concerns} \citeSLR{Corbalan-Navarro20224375} \citeSLR{hosseini2012energy} \citeSLR{park2016hicap} \\ \hline
\multicolumn{1}{|p{0.02\linewidth}|}{D}  & \citeSLR{6471008} & \citeSLR{Yan201813} \citeSLR{9591818} & & \citeSLR{hosseini2012energy} \\ \hline
\multicolumn{1}{|p{0.02\linewidth}|}{A} & & & & \\  \hline
\end{tabular}
\end{table}

\subsection{RQ4. What other video game aspects should be considered to develop Green Video Games?}

In this section, we outline different aspects of video games that are largely neglected in Green Computing. Specifically, we focus on video game artifacts, genres, and content.

\subsubsection{Artifacts}
\hfill

Among the main artifacts of video games, game engines and physics engines are known for their complexity and consumption of resources. Game engines (e.g., Unity or Unreal) are software tools for developing video games that use libraries or support programs instead of creating everything from the ground up. Physics engines are software components that simulate the motion of objects based on parameters such as mass, friction, or gravity.

One way to develop greener video games could be to improve energy efficiency during their development by improving the energy efficiency of the game engines and physics engines used to develop video games. However, according to our survey, only four primary studies take these artifacts into account.

The authors in \citeSLR{gu2006games} presents a novel characterization of the workload of computer games based on a game engine. This same authors present in \citeSLR{gu2008power} a novel dynamic voltage scaling scheme that is specifically directed towards 3D graphics-intensive interactive gaming applications running on battery operated portable devices. In both papers, the importance of the game engine (Quake II) and the physics engine are highlighted. They used these engines to evaluate the proposed algorithms. However, neither of these works seeks to improve the energy efficiency of the game engine. 

In contrast, the other two primary studies focus on improving the energy efficiency of physics engines. The work in \citeSLR{bose2012physics} demonstrates that physics processing can be optimized using a reconfigurable processor. In addition, \citeSLR{Toupas201988} considers the use of FPGAs to accelerate certain demanding components of the physics simulation pipeline aiming to provide better performing solutions at a significantly lower energy cost.

In terms of the motivation and the device, the two studies that consider game engines (\citeSLR{gu2006games} and \citeSLR{gu2008power}) focus on battery life in portable devices. In contrast, the two studies that consider physics engines (\citeSLR{bose2012physics} and \citeSLR{Toupas201988}) focus on the Trade-off between energy consumption and application performance. In terms of the layer, the four studies focus on the hardware layer.

\subsubsection{Genres}
\hfill

Today, there are many genres in video games, and each game has different requirements. For example, an offline single-player game does not require the same infrastructure and energy as MMOG. From the primary studies, we have identified the following genres:

\begin{itemize}
 \item \textbf{First-Person Shooter (FPS)} is a gun-based video game in which the player observes the scenario from the protagonist's point of view, i.e., the user carries a weapon and sees the protagonist's perspective in first-person. For the user to achieve good immersion, 3D graphics are essential in this genre. In addition, some FPS video games may be played by multiple users online. Battlefield 2042 or Far Cry 6 are examples of FPS video games.
 \item \textbf{Massively Multiplayer Online Game (MMOG)} is an online video game that is played by a large number of users through the same server. World~of~Warcraft or Destiny~2 are examples of MMOG.
 \item \textbf{Real-Time Strategy (RTS)} is a strategy video game where all of the users play simultaneously (in real-time) instead of playing in turns. The games are usually fast (maximum one hour) and dynamic. Some games of this genre are played online and/or have a multiplayer mode. Crusader~Kings~III or Age~of~Empires~IV are examples of RTS video games.
 \item \textbf{Board games} are video games based on board games. Stacklands or UNO are examples of Board games.
\end{itemize}

\begin{figure*}
\centering
\includegraphics[width=.45\textwidth]{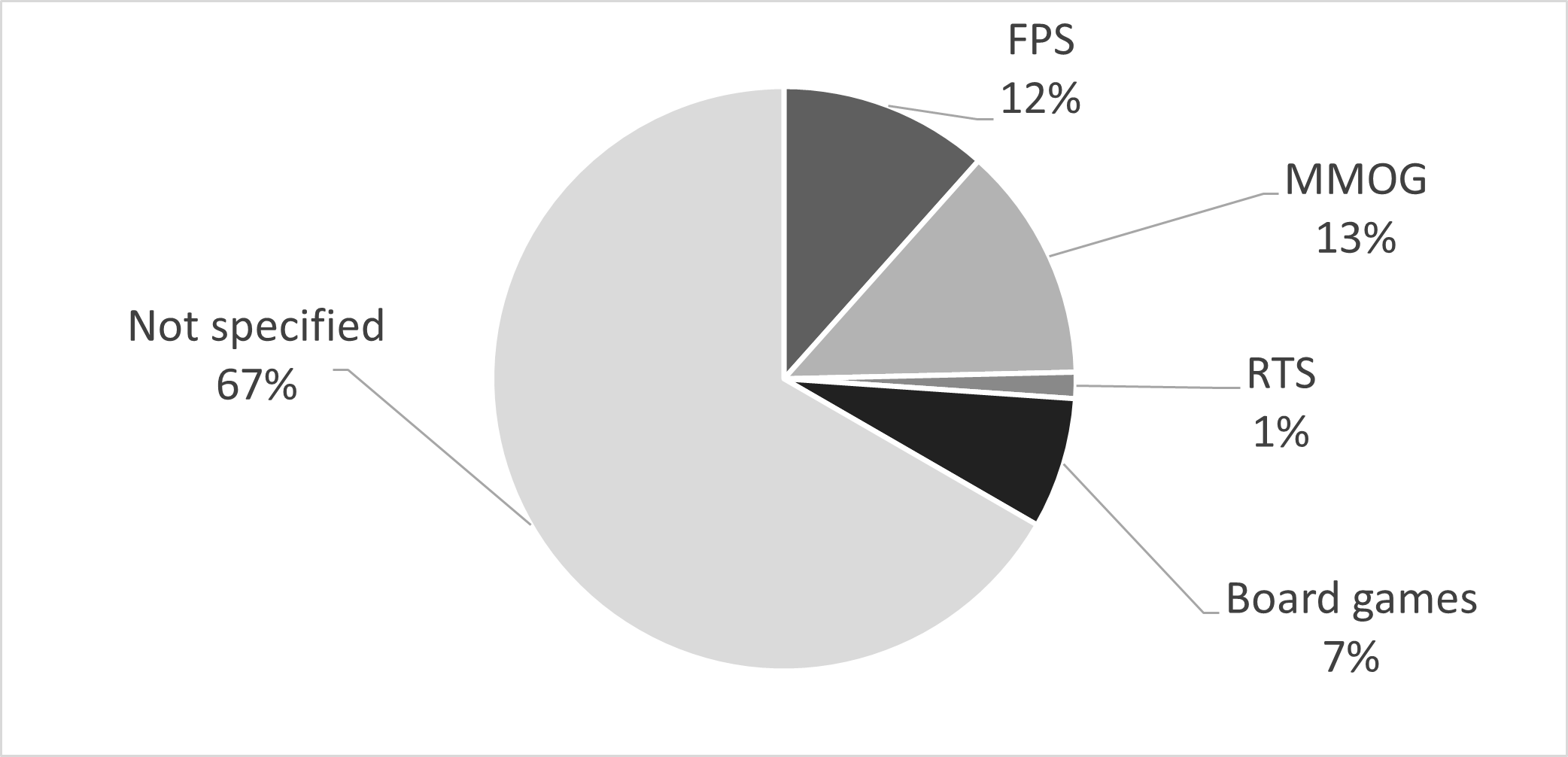}
\caption{Distribution of the primary studies based on genres of video games}
\label{Fig_RQ4_1}
\end{figure*}

Fig. \ref{Fig_RQ4_1} shows the genres of the video games used in the primary studies. Most of the primary studies do not mention the genre of the video games used to evaluate their work. FPS and MMOG are the genres that are most used in the primary studies (12\% and 13\% respectively). Board games are used in 7\% of the primary studies, and only 1\% of the primary studies use RTS video games. 

In terms of the motivation (Table \ref{tab:RQ4_1}), most of the primary studies using FPS and MMOG video games focus on Battery life or the Trade-off between energy consumption and other aspects. There is only one study that uses RTS video games, and it also focuses on Battery life. However, most of the studies using Board games focus on Software and development optimization. With regard to Modelling and measurement of energy consumption, all of the studies use MMOG. Moreover, the studies that focus on General motivation do not specify the genre because they are not using a video game to evaluate their work.

\begin{table}[]
\centering
\renewcommand{\arraystretch}{1.2}
\caption{Genres of the video games used in primary studies over motivation (1-Battery life, 2-Data center design, 3-Modelling and measurement of energy consumption, 4-General, 5-Software and development optimization, and 6-Trade-off between energy consumption and other aspects).}
\label{tab:RQ4_1}
\begin{tabular}{p{0.01\linewidth}|p{0.20\linewidth}|p{0.23\linewidth}|p{0.20\linewidth}|p{0.20\linewidth}|p{0.20\linewidth}}
 \cline{2-5}
 & FPS & MMOG & RTS & Board games \\ \hline
\multicolumn{1}{|p{0.01\linewidth}|}{1} & \citeSLR{Muhuri20182311} \citeSLR{gu2008control} \citeSLR{thirugnanam2012dynamic} \citeSLR{anand2009game} \citeSLR{gu2006games} & \citeSLR{8057058} \citeSLR{chen2015user} \citeSLR{pinto2017energy} & \citeSLR{5679572} & \\ \hline
\multicolumn{1}{|p{0.01\linewidth}|}{2} & & \citeSLR{Dhib20222119} \citeSLR{nae2008efficient} & & \\ \hline
\multicolumn{1}{|p{0.01\linewidth}|}{3} & \citeSLR{chen2018user} & \citeSLR{6471008} & & \\ \hline
\multicolumn{1}{|p{0.01\linewidth}|}{4} & & & & \\ \hline
\multicolumn{1}{|p{0.01\linewidth}|}{5} & & & & \citeSLR{LeBorgne201263} \citeSLR{Rodriguez202218704} \citeSLR{7380525} \\ \hline
\multicolumn{1}{|p{0.01\linewidth}|}{6} & \citeSLR{Suriano2020136} \citeSLR{8566409} & \citeSLR{Kavalionak2015301} \citeSLR{7037666} \citeSLR{ahmadi2014game} & & \citeSLR{Olivito201564} \citeSLR{lee2016exploiting} \\ \hline
\end{tabular}
\end{table}

In terms of the device (Table \ref{tab:RQ4_2}), FPS video games are used overall by studies that focus on handheld and mobile devices. In contrast, MMOG video games are used overall by studies that focus on cloud computing and handheld devices. Cloud computing is a valuable resource for MMOG games since they are played on a server. Therefore, it is logical for most of the studies on cloud computing to be evaluated using MMOG games.

The study that uses RTS video games \citeSLR{5679572} focuses on mobile devices. Finally, Board games are used to evaluate studies that focus on multiple devices, such as computers, mobile devices, or wireless sensors.

\begin{table}[]
\centering
\renewcommand{\arraystretch}{1.2}
\caption{Genres of the video games used in primary studies over devices (D1-Cloud computing, D2-Computers, D3-Edge devices, D4-Gaming devices, D5-Handheld devices, D6-Mobile devices, D7-Portable devices, D8-Reconfigurable MPSoCs, D9-Smart devices, D10-Wearable devices, D11-Wireless sensors, and D12-Not specified).}
\label{tab:RQ4_2}
\begin{tabular}{p{0.03\linewidth}|p{0.20\linewidth}|p{0.23\linewidth}|p{0.20\linewidth}|p{0.20\linewidth}|p{0.20\linewidth}}
 \cline{2-5}
 & FPS & MMOG & RTS & Board games \\ \hline
\multicolumn{1}{|p{0.03\linewidth}|}{D1} & & \citeSLR{Kavalionak2015301} \citeSLR{7037666} \citeSLR{Dhib20222119} \citeSLR{ahmadi2014game} \citeSLR{nae2008efficient} & & \\\hline
\multicolumn{1}{|p{0.03\linewidth}|}{D2} & & & & \citeSLR{7380525} \\\hline
\multicolumn{1}{|p{0.03\linewidth}|}{D3} & & & & \\\hline
\multicolumn{1}{|p{0.03\linewidth}|}{D4} & & & & \\\hline
\multicolumn{1}{|p{0.03\linewidth}|}{D5} & & & & \\\hline
\multicolumn{1}{|p{0.03\linewidth}|}{D6} & \citeSLR{chen2018user} \citeSLR{8566409} \citeSLR{thirugnanam2012dynamic} \citeSLR{anand2009game} & \citeSLR{8057058} \citeSLR{chen2015user} \citeSLR{pinto2017energy} & \citeSLR{5679572} & \citeSLR{Olivito201564} \citeSLR{lee2016exploiting} \\\hline
\multicolumn{1}{|p{0.03\linewidth}|}{D7} & \citeSLR{Muhuri20182311} \citeSLR{gu2008control} \citeSLR{gu2006games} & \citeSLR{6471008} & & \\\hline
\multicolumn{1}{|p{0.03\linewidth}|}{D8} & \citeSLR{Suriano2020136} & & & \\\hline
\multicolumn{1}{|p{0.03\linewidth}|}{D9} & & & & \\\hline
\multicolumn{1}{|p{0.03\linewidth}|}{D10} & & \citeSLR{pinto2017energy} & & \\\hline
\multicolumn{1}{|p{0.03\linewidth}|}{D11} & & & & \citeSLR{LeBorgne201263} \\\hline
\multicolumn{1}{|p{0.03\linewidth}|}{D12} & & & & \citeSLR{Rodriguez202218704} \\\hline
\end{tabular}
\end{table}

In terms of the layer (Table \ref{tab:RQ4_3}), most of the studies that use FPS video games present solutions for the hardware layer. As previously mentioned, in order to achieve good immersion of the user in FPS video games, 3D graphics are essential in this genre. Therefore, the consumption of the CPU and GPU rendering 3D is a point for improving sustainability in video games.

In contrast, most of the studies that use MMOG video games present solutions for the network layer. In this case, the server is an essential component for this genre: how the users are connected simultaneously at the same server, how the data of the video game is transferred, what the latency is, etc. Therefore, saving energy on the network layer may be interesting for MMOG video games.

Finally, most studies that use RTS and Board games present solutions for the software layer. RTS video games and especially Board games do not require complex computations through the CPU or GPU. In addition, many of these video games are played offline, so they do not require complex or power-hungry components at the network layer. However, the efficiency of these video games can be improved by using more efficient algorithms.

\begin{table}[]
\centering
\renewcommand{\arraystretch}{1.2}
\caption{Genres of the video games used in primary studies over layers (H-Hardware, N-Network, S-Software, D-Design, and A-Art).}
\label{tab:RQ4_3}
\begin{tabular}{p{0.02\linewidth}|p{0.20\linewidth}|p{0.23\linewidth}|p{0.20\linewidth}|p{0.20\linewidth}|p{0.20\linewidth}}
 \cline{2-5}
 & FPS & MMOG & RTS & Board games \\ \hline
\multicolumn{1}{|p{0.02\linewidth}|}{H} & \citeSLR{Muhuri20182311} \citeSLR{Suriano2020136} \citeSLR{chen2018user} \citeSLR{gu2008control} \citeSLR{gu2006games} & \citeSLR{6471008} \citeSLR{chen2015user} & & \citeSLR{Olivito201564} \\ \hline
\multicolumn{1}{|p{0.02\linewidth}|}{N} & \citeSLR{8566409} & \citeSLR{Kavalionak2015301} \citeSLR{7037666} \citeSLR{8057058} \citeSLR{Dhib20222119} \citeSLR{ahmadi2014game} \citeSLR{nae2008efficient} & & \citeSLR{lee2016exploiting} \\ \hline
\multicolumn{1}{|p{0.02\linewidth}|}{S} & \citeSLR{Suriano2020136} \citeSLR{thirugnanam2012dynamic} \citeSLR{anand2009game} & \citeSLR{pinto2017energy} & \citeSLR{5679572} & \citeSLR{LeBorgne201263} \citeSLR{Olivito201564} \citeSLR{Rodriguez202218704} \citeSLR{7380525} \\ \hline
\multicolumn{1}{|p{0.02\linewidth}|}{D} & & \citeSLR{6471008} & & \\ \hline
\multicolumn{1}{|p{0.02\linewidth}|}{A} & & & & \\ \hline
\end{tabular}
\end{table}

\subsubsection{Content}
\hfill \break
The content consists of the narrative, gameplay, technical, audio, and visual components of a video game, from the storyboards to characters, scenarios, music, etc. From the primary studies, we have identified four types of content: maps, lights, textures, and video broadcasting or encoding.

In video games, maps are used as the main user interface, as a reference guide, or both. One primary study focuses on maps. Specifically, \citeSLR{Nery201997} tackled the consumption of pathfinding algorithms. Its experimental results are presented including performance, circuit-area, and energy consumption results for a set of artificial and commercial (game) maps.

Lights are not only important for visualizing the depth of a scene or an object. They are also used to provide information to help players or to evoke moods. For example, in horror video games, there are many different shadows for hiding or provoking a sense of tension in the user. Two of the primary studies tackled the consumption of video games and virtual environments that offer immersive experiences. The work in \citeSLR{Yan201813} shifts the default fixed full brightness in video games to a dark adaptation based dynamically scaled brightness. The work in \citeSLR{hosseini2012energy} proposes two new energy-aware adaptation schemes that can be employed in 3D graphics applications. One of these two schemes is based on lighting limitation.

The other scheme is based on textural transformation. Textures are also content in video games. Textures are the graphical skins that render the objects within a video game. Textures require memory space for storage and bandwidth for transfer to the graphics card for rendering. To store and transfer the textures for rendering, video games make use of the GPU and the bandwidth, respectively. Therefore, \citeSLR{hosseini2012energy} focuses on energy saving with acceptable sacrifices to a user's visual experience, specifically transforming to the textures.

Finally, five of the primary studies consider video broadcasting or encoding. Three of these primary studies focus on different strategies for cloud gaming. The work in \citeSLR{Gharsallaoui20171072} proposes a new strategy that takes into consideration the quality of service parameters and the energy-consumption of the decoder. The work in \citeSLR{Ghergulescu2014} proposes an energy-aware adaptive multimedia game-based e-learning framework that builds on top of the idea to render the game on the server side and stream a recording of it to the player's device over the Internet. The work in \citeSLR{ahmadi2014game} introduces the concept of game attention model (GAM), which reduces the bit rate of the streaming video more efficiently. 

In addition, the other two primary studies belong to the same authors and focus on broadcast game playthroughs in real time. In \citeSLR{8057058}, the authors present a novel interaction-aware optimization framework to improve the energy utilization and stream quality for mobile gamecasting. This framework is based on their measurements revealing that there are strong associations between the users' touch interactions and the viewers' gazing patterns. This study was then extended by the authors in \citeSLR{zhang2017exploring}, where they include more details about the framework and the evaluation and include a further discussion.

In terms of the motivation, these primary studies focus on Battery life or the Trade-off between energy consumption and other aspects. In fact, all of the primary studies that tackled Battery life (i.e., \citeSLR{Yan201813}, \citeSLR{hosseini2012energy}, \citeSLR{Ghergulescu2014}, \citeSLR{zhang2017exploring}, and \citeSLR{8057058}) focus on mobile devices. In contrast, two out of the three primary studies which tackled the Trade-off between energy consumption and other aspects focus on cloud computing (i.e., \citeSLR{Gharsallaoui20171072} and \citeSLR{ahmadi2014game}), and the other one focuses on gaming devices (i.e., \citeSLR{Nery201997}).

In terms of the layer, most of these primary studies present solutions on the network layer (i.e., \citeSLR{Ghergulescu2014}, \citeSLR{zhang2017exploring}, \citeSLR{8057058}, \citeSLR{ahmadi2014game}). Nevertheless, there are studies on the rest of the layers, except for the art layer. The solution in \citeSLR{Nery201997} relies on the hardware layer, the solution in	\citeSLR{Gharsallaoui20171072} relies on the software layer, and the solution in \citeSLR{Yan201813} relies on the design layer. The solution in \citeSLR{hosseini2012energy} relies in both the software and the design layer.


\section{Discussion}
\label{section:discussion}

Through a thorough analysis of the results obtained by our SLR, we have raised a series of discussion points regarding the state of the art on Green Video Games. The following subsections bring forth such a discussion.

\subsection{Limitation of the Video Game issues}

In Green Video Games, many works address issues that are related to the energy sustainability of video games; however, not always this energy sustainability is not always the main motivation. For example, \citeSLR{Muhuri20182311} highlights battery life as one of the drawbacks to user satisfaction. Therefore, by consuming less power, the battery will last longer, which will in turn have a positive impact on user satisfaction. The main motivation is not to develop greener video games, but to improve battery life to the satisfaction of users. There is indirection between the sustainability solution proposed by the study and the motivation of the study. 

Moreover, there is no common way to define an issue in Green Video Games. The motivation and solution of one work can be interchanged depending on the perspective of the research. For example, the motivation of \citeSLR{7351775} considers that the battery in mobile devices is a critical reason for controlling consumption, so it proposes an energy efficiency evaluation mechanism that allows to control of the energy consumed by the CPU-GPU. Therefore, according to this motivation, the study would be classified as a Battery life issue, but according to its solution, it would be classified as a Modelling and measurement of energy consumption issue. To avoid this ambiguity in the issue definition, we propose always focusing on motivation.

In the case of solutions, there is a wide range of different solutions, so it is difficult to identify commonalities for them. From the primary studies, we identified two commonalities: devices and layers. Device refers to the electronic device on which end-users play a video game, so the solution should be evaluated on this device. Layer refers to the video game layer (i.e., hardware, network, software, etc.) on which components are affected or enhanced by the solution. Both commonalities may be improved.

With regard to devices, there are a lot of similar categories: wearable devices, smart devices, portable devices, handheld devices, or mobile devices. When thinking about the definitions, there are small differences that characterize each one of these categories. However, why are the differences relevant for the studies and why do the studies focus more on mobile devices instead of portable devices or handheld devices? As yet, we have not been able to answer these questions. From the primary studies, there are no clear reasons for improving the battery life on a mobile device instead of improving the battery life on a portable device or a handheld device. A possible reason for this breach on the device in study might be that the mobile games revenue accounted for 52\% (\$93.2Bn) of the global market in 2021 while the console games revenue accounted for 28\% (\$50.4Bn)~\cite{wijman2021games}. We also have to take into account that most consoles are not portable or handheld devices, so only a fraction of this percentage belongs to these devices.

With regard to layers, they were identified by professional video game developers to cover all of the relevant aspects of video game creation. However, not all of these layers have been considered for improving the sustainability of video games. Are these layers not relevant to creating green video games? From our point of view, all layers can be used to improve the sustainability of a video game, but researchers have not yet noticed the potential of video game-exclusive layers. This is a topic that will be discussed in depth later.

To sum up, in Green Video Games, we found two main limitations to identifying issues: 1) the indirection between the sustainability solution and the motivation of the study, and 2) the ambiguity in identifying an issue by its motivation or its solution. These problems make it difficult to find related works and compare results with other works. To solve these problems, we propose defining Green Video Game issues based on motivation, device, and layer. Although this definition has limitations, it can be a starting point that will be improved depending on how the Green Video Game area evolves.

\subsection{From Green Computing to Green Video Games}

Concern for energy efficiency in video games is a relatively new field of research. According to the demographic information extracted, interest has been growing since around 2007 to the present day. However, although it is not an extensive field of research, researchers have already proposed many solutions to improve energy sustainability at the hardware, network, and software levels.

Moreover, the first three layers of the issues (i.e., hardware, software, and network) have been addressed at the same time since the beginning of the research field. This is a fact that we consider surprising and completely different from our expectations. Perhaps because of our personal experience or because of our engineering careers, we assumed that the issues at the lowest layer (hardware) would be investigated first and then the research would progress to the higher layers when the results were consolidated.

However, the results obtained in this work have proven us wrong, which has given us food for thought. \textit{Is this a consequence of the diversity of the venues?} or \textit{Could there be other reasons that we have not yet considered?}

Of the 69 primary studies, 31 have been published in conferences and 25 in journals. When comparing conferences, we have seen that there are 26 different conferences and 21 different journals. In other words, the ratio between the number of conferences and the number of studies is 0.84, and the ratio between the number of journals and the number of studies is 0.84. Eighty-four percent of the studies have been published in different venues. At first glance, there appear to be no conferences or journals of reference or prominence in this area of research.

Some conferences or journals focus more on hardware (e.g., IEEE International Conference on Consumer Electronics), others on networks (e.g., International Wireless Communications and Mobile Computing Conference), and still yet others on software (e.g., International Conference on Software and Information Engineering). However, there are also conferences and journals whose topics of interest are more generic, such as the Design Automation Conference or IEEE Systems Journal.

Considering the above data, none of the venues focus on Green Video Games, and most have only published one paper in this field. This leads us to think that engineers researching other Green Computing problems have started to show interest in transferring their findings to the field of video games. 

Green Computing is a research area whose findings and techniques in hardware, networks, and software are already consolidated. Therefore, transferring these findings and techniques to the field of video games, where there is high energy consumption, would be a natural step from Green Computing. It would also explain why most primary studies do not take into account other layers of video games, such as design or art.

\subsection{Energy-awareness}

Based on our observations and the results obtained, we believe that there is a lack of environmental awareness in video games. Everything points to the fact that energy sustainability in video games is a secondary aspect when designing, developing, and maintaining video games.

As we commented in the previous section, the main motivation of the studies is not to improve the energy sustainability of video games. There is indirection between the sustainability solution and the motivation of the study. For example, many studies focus on video games installed on mobile devices and their main motivation is to extend battery life to improve user satisfaction. A prematurely depleted battery has a negative impact on the user experience and can negatively affect the rating of the video game.

In addition to the impact on battery life, the progressive increase in the complexity of video games (i.e., graphics, gameplay, interaction with other users) also affects the number and cost of resources required to deploy the video game, especially in the increasingly common MMOG. In these games, a large number of users connect to the same server to enjoy online video games. MMOGs require resources such as communication infrastructures and, above all, data centers configured in the cloud.

It seems that the core around which all motivations revolve is the developer's profit. Improving some aspects of the video game means that there is a profit for the developers. For example, if users are happy they keep playing our game, so we have to solve the problems that frustrate users (battery consumption, latency, CPU or GPU consumption, etc.). If we improve the design of the data center we spend less money on the energy consumed. 

What we want to emphasize is that sustainability should not be a side effect that we benefit from when it happens. Rather, it should be one of the concerns or objectives during the development of the video game. This is something that is not yet being achieved according to the results obtained in this survey. However, we encourage researchers to consider all of the advantages of sustainable development for video games.

\subsection{Trends and opportunities}

Thanks to the survey carried out, we have been able to observe certain trends and identify some points as future lines of research. These trends and future lines are included below.

\begin{enumerate}
\item We have noticed that some of the issues in Green Computing (i.e., Product longevity, material recycling, and telecommuting) are not yet being taken into account for Green Video Games. This leaves some open questions: \textit{ Do video games have an expiration date? Can consoles or devices be reused? How should gaming devices be recycled? Can cloud computing be considered a way to minimize the material to be recycled? Can telecommuting be used to improve the energy sustainability of video games?}.

\item We have observed that few studies propose solutions from the design layer and none of the studies propose solutions from the art layer. Although not much attention is being paid to these layers, we believe they could be considered an opportunity. At the design layer, developers could take advantage of elements designed to minimize power consumption. For example, in a video game, a map could be designed with two main scenarios connected by a corridor. As the player travels down that corridor, the main scenario is loaded in that user's direction. This way, if the user heads toward the second main scenario, it will load, or vice versa. Instead of loading the entire game at the beginning, the main scenarios are loaded based on the user's location. In addition, the game does not have to wait for the next scenario to load, because while the loading occurs, the user is entertained by walking the corridor. In the end, the energy and resources required to run a single main scenario should be less than that required to run the entire game. 

In fact, there are video games that employ the same or similar strategies. However, strategies are employed without addressing how they influence the energy consumption of the video game. Furthermore, they are often not documented or reported, so the research community cannot benefit from the ecological advances made. Therefore, there is still a need to assess and report on these types of strategies from an ecological perspective.

Similarly, in the art layer, developers could consider how to save energy by taking advantage of the artistic components of video games. For example, the rhythm of a song may encourage the player to walk more slowly or dialogues may make the player stop to read its content, which could provide the processor with a few moments of relief or cede resources to other tasks. Also, simpler textures and the usage of 2D meshes instead of 3D meshes would reduce the resources needed by the GPU while maintaining an aesthetic that is attractive for users. There are examples of successful commercial video games that use these techniques, like Void Bastards and Cult of the Lamb.

\item There are very few works that take into account artifacts, such as the game engine or physics engine, despite their high energy consumption. The video game genre is also barely considered, even though the energy consumption is not the same for a single or multiplayer game played locally or online. 

Furthermore, only eight of the primary studies address the development of Green Video Games with specific content in mind (i.e., textures, lights, maps, etc.). For example, complex textures not only add to the diversity of colors, but they make the application of lighting effects, which have to be computed for all 3D textures within the world, more expensive since this involves modifying a large number of memory objects and stresses the memory system \citeSLR{hosseini2012energy}. A decade has passed since the publication of this paper, but it is the only paper that considers textures as a way to reduce energy consumption.

Everything seems to indicate that these three aspects (artifacts, genre, and content) can be leveraged to improve the effectiveness of the video game. However, researchers should think from the point of view of video games, rather than seeing them as just another application for their devices. Video games have their artifacts, genres, content, and even other specific aspects that we have not considered in this study. The question is how researchers can take advantage of these aspects or others to develop greener video games.

\item We have noted the lack of threats to validity, limitations, and /or future work. None of the studies included threats to validity. The few included limitations are too specific and there are no relation among them. Therefore, the main limitation of Green Video Games today may be the youthfulness of this area of research. Green Video Game research would improve rapidly if researchers would compare and discuss the results of their studies with each other. This seems difficult in the current state: 1) Green Video Games studies are published in a wide diversity of venues, making it difficult to find related publications or to identify all researchers working on related articles; 2) There is ambiguity and redirection in the description of issues, so it also complicates finding related articles; 3) Most studies do not report data sets, devices, techniques, threats to validity, or limitations, making it difficult to compare results fairly. Nevertheless, we hope that the overview provided by this survey will help somewhat to minimize these problems and encourage researchers to report their studies more fully.
\end{enumerate}

\section{Threats to validity}
\label{section:threats}

\subsection{Threats to validity}

In this section, we use the classification of threats to validity of~\cite{VAL:Wohlin12} and the study of threats to validity of systematic literature reviews in \cite{zhou2016map} to acknowledge the limitations of our survey.

\textbf{Construct validity:} This aspect of validity reflects the extent to which the operational measures that are studied represent what the researchers have in mind. Examples of issues are whether the concepts are defined clearly enough before measurements are defined and the interaction of different treatments when people are involved in more than one study~\cite{wohlin2003empirical}. 

\begin{itemize}
 \item To avoid the threat of non-specification of survey setting and sufficient details, we describe the search string, the digital sources, the inclusion and exclusion criteria, and the data extraction in detail.

 \item To avoid the threat of inappropriate or incomplete search terms in automatic search, we used the steps suggested by Kitchenham and Charters~\cite{kitchenham2007guidelines}. First, we used the PICO (Population, Intervention, Comparison, and Outcomes) criteria to derive the major terms from the RQs. Then, we identified synonyms for these terms. Finally, we verified the search terms that other relevant surveys used in their search strings.

 \item To avoid the threat of incorrect search method, our search strategy is based on the guidelines provided in ~\cite{kitchenham2007guidelines,kitchenham2013systematic,kitchenham2015evidence}.

 \item To avoid the threat of inappropriate exclusion and inclusion criteria, the exclusion criteria were defined according to criteria commonly used in surveys. Moreover, to minimize this threat, the inclusion criteria were defined by domain experts in order to reduce the probability of not including studies whose conclusion could be relevant for our research field.

 \item To minimize the threat of inappropriate research questions, specific sessions were held with all of the co-authors to discuss and assess which research questions were most appropriate in order to provide an overview of Green Video Games that would be useful to both researchers and practitioners.
 
\end{itemize}

\textbf{Internal Validity:} This aspect of validity is of concern when causal relations are examined. There is a risk that the factor being investigated may be affected by other neglected factors. In this work, the outcomes could be affected by how the primary studies are selected. 

\begin{itemize}
 \item To minimize the threat of misclassification of primary studies, the studies were classified by two authors of the paper independently. Then, their results were compared. In case of disagreement, the first and second authors held a discussion in order to reach a consensus.

 \item To avoid the threat of primary study duplication, duplication was included as an exclusion criterion.
 
 \item To minimize the threat of bias in study selection, the selection was conducted by two authors of the paper. Therefore, the selection process was double-checked. Moreover, they strictly followed the search strategy defined for the selection.

 \item To minimize the threat of bias in data extraction, similarly to the previous threat, the extraction was conducted by two authors of the paper. Moreover, the features for data extraction was identified before starting the extraction process.
 
\end{itemize}

\textbf{External Validity:} This aspect of validity is concerned with to what extent it is possible to generalize the findings, and to what extent the findings are of relevance for other cases. There is a risk that the papers recovered by the search are not representative of the target population. 

\begin{itemize}
 \item To minimize the threat of incomplete research information or conclusions by primary studies, we used snowballing to avoid missing relevant studies.

 \item To avoid the threat of restricted time span, we did not include any temporal restriction in our search method except for the date on which the search was performed in the digital sources.
\end{itemize}

\textbf{Conclusion validity:} This aspect is concerned with to what extent the process can be replicated with the same results. Researchers may influence the result by looking for a specific outcome.

\begin{itemize}
 \item To minimize the threat of subjective interpretation of the extracted data, we clearly separated the results from the discussion. The results show extracted data for answering RQs. The discussion provides different points regarding the state of Green Video Games according to the author's interpretation of the results.

 \item To avoid the threat of replication of the study, the documents of the whole search process (step by step) have been made publicly available. 
\end{itemize}

\section{Conclusions}
\label{section:conclusions}

Today, there is a huge number of video game players that use everything from game consoles that are created specifically for playing video games to computers and mobiles. The combination of the large number of users with the high computational requirements of modern video games has motivated research on Green Video Games. However, this is recent research where publications are scattered, so it is difficult to find related works, compare results, or even determine trends or future lines for research work. For this reason, the main objective of this survey is to provide an overview of Green Video Games publications. This article also aims to serve to identify recommendations for practitioners, outstanding research challenges, and several potential avenues of future research for Green Video Games.

In this paper, the first survey in this field of study, we have systematically reviewed the works in Green Video Games. We have reviewed a total of 2,637 papers, selecting 69 studies for further analysis. The results of this survey bring light to the current state of the art. Through a detailed analysis of the results, we propose a new way to define the Green Video Game issues based on the motivation of the studies and the device and the layer on which the solutions are evaluated. Then, based on the Green Video Games issues, we analyzed the different applied techniques, the limitations and levels of evidence, and some aspects that are specific to video games. 

The overview provided by this paper can be useful for practitioners and researchers alike when studying the current state of the field. Additionally, through the recommendations provided, along with the discussion of the results, our survey has the potential to produce a positive impact in the Video Green Computing area highlighting that video games have specific aspects that can be leveraged to develop even greener video games. Our study focuses on enhancing the understanding of those works that improve the sustainability of video games. This paper could be complemented by analyzing related fields of research that are left open as interesting opportunities for future literature exploration, e.g., in the form of reviews that tackle the use of video games to raise people’s awareness of environmental impact or to teach them ecological practices for a more sustainable environment.

\section*{Acknowledgements}
This work was supported in part by the Ministry of Economy and Competitiveness
(MINECO) through the Spanish National R+D+i Plan and ERDF funds under the Project VARIATIVA under Grant PID2021-128695OB-I00, and in part by the Gobierno de Arag\'on (Spain) (Research Group S05\_20D).

\bibliographystyle{plain}
\bibliography{bibliography}

\nociteSLR{*}
\bibliographystyleSLR{plain}
\bibliographySLR{selectedPapersVF}

\end{document}